\documentclass{aa}  
\usepackage{natbib}
\bibpunct{(}{)}{;}{a}{}{,} 
\usepackage{amsmath}
\usepackage{lineno}
\usepackage[colorlinks=true,linkcolor=blue,citecolor=blue]{hyperref}

\linenumbers

\begin{document}
\title{Interior dynamics of super-Earth 55 Cancri e}

\title{
Interior dynamics of super-Earth 55 Cancri e
}

\author{Tobias G. Meier \inst{1,2} \and Dan J. Bower \inst{1} \and Tim Lichtenberg \inst{3} \and Mark Hammond \inst{2} \and Paul J. Tackley \inst{4}}
\institute{Center for Space and Habitability, University of Bern, Gesellschaftsstrasse 6, 3012 Bern, Switzerland\\ \email{tobias.meier@physics.ox.ac.uk} \and 
Atmospheric, Oceanic and Planetary Physics, Department of Physics, University of Oxford, Parks Road, Oxford OX1 3PU, United Kingdom \and
Kapteyn Astronomical Institute, University of Groningen, P.O. Box 800, 9700 AV Groningen, The Netherlands \and 
Institute of Geophysics, Department of Earth Sciences, ETH Zurich, Sonneggstrasse 5, Zurich 8092, Switzerland
}

\date{Received 22 May, 2023; accepted 24 July, 2023}
\abstract
{
The ultra-short-period super-Earth 55 Cancri e has a measured radius of $1.8$ Earth radii. Previous thermal phase curve observations suggest a strong temperature contrast between the dayside and nightside of around $1000$\,K with the hottest point shifted $41\pm12$ degrees east from the substellar point, indicating some degree of heat circulation. The dayside (and potentially even the nightside) is hot enough to harbour a magma ocean. We use results from general circulation models (GCMs) of atmospheres to constrain the surface temperature contrasts.

There is still a large uncertainty on the vigour and style of mantle convection in super-Earths, especially those that experience stellar irradiation large enough to harbour a magma ocean. In this work, we aim to constrain the mantle dynamics of the tidally locked lava world 55 Cancri e.
Using the surface temperature contrasts as boundary condition, we model the mantle flow of 55 Cancri e using 2D mantle convection simulations and investigate how the convection regimes are affected by the different climate models.

We find that large super-plumes form on the dayside if that hemisphere is covered by a magma ocean and the nightside remains solid or only partially molten. Cold material descends into the deep interior on the nightside, but no strong downwellings form. In some cases, the super-plume also moves several tens of degrees towards the terminator.  
A convective regime where the upwelling is preferentially on the dayside might lead to preferential outgassing on that hemisphere which could lead to the build-up of atmospheric species that could be chemically distinct from the nightside.}

\maketitle 
\nolinenumbers
\section{Introduction} \label{sec:intro}
Many discovered exoplanets fall into the category of so-called super-Earths. These are rocky exoplanets that are more massive than Earth, but less massive than the ice giants \citep[e.g.][]{Haghighipour2011}. These planets are absent in our solar system and attention has therefore been drawn towards understanding their formation mechanisms \citep[e.g.][]{Raymond2008, Schlichting2018} and potential habitability \citep[e.g.][]{Bloh2007, Seager2013, Madhusudhan2021, MolLous2022}. A key component in assessing a planet’s potential for habitability, is whether it is able to sustain a long-term volatile cycle between the atmosphere and its interior \citep[e.g.][]{Dehant2019}. For this, it is crucial to better understand the possible tectonic regimes that super-Earths might exhibit.  While plate tectonics can provide an efficient mechanism to regulate a planet’s climate \citep[][]{Foley2015, Oosterloo2021}, planets with an immobile lid (so-called stagnant lid planets) could also have temperate climates \citep[e.g.][]{Tosi2017, Foley2019, Unterborn2022}. Several studies have found that the outgassing rates on stagnant lid planets depend on various factors such as mass \citep[e.g.][]{Noack2017, Dorn2018a}, bulk composition \citep[e.g.][]{Spaargaren2020} and redox state \citep[e.g.][]{Guimond2021,  Liggins2022, Baumeister2023}.
Because the vigour and style of mantle convection in super-Earths is still largely unknown, here we use global mantle convection simulations to investigate the interior dynamics of super-Earth 55 Cancri e \citep{McArthur2004}, for which observational data suggest that its dayside (and potentially nightside) is covered by a magma ocean. 

55 Cancri e has a measured radius of $1.8$ Earth radii and its thermal phase curve indicates a nightside temperature around $1400$\,K with a significant shift of the hottest point ($2700$\,K) located $41\pm12$ degrees east from the substellar point \citep{Demory2016a} indicating heat redistribution. 
\citet{Hammond2017} have run general circulation models (GCMs) for different types of atmospheres. Their best-fitting model consists of an atmosphere with a $90$\%-$10$\% mixture of H$_2$ and N$_2$ with a mean-molecular weight $\mu=4.6$\,gmol$^{-1}$, optical depth $\tau=4.0$, and surface pressure $p_s=5$\,bar, which produces a significant hot spot and temperature contrast, albeit not as large as those observed.  

Because of the high solar irradiation, the planet is most likely covered by a magma ocean, placing it into the category of lava planets \citep{Chao2021,Lichtenberg2023}. Depending on the amount of heat circulation between the dayside and nightside, a hemispheric magma ocean could form with a molten dayside and a cold, rocky nightside \citep{Leger2011,Pierrehumbert2019}. Subsequent outgassing of volatiles (e.g., CO, CO$_2$, H$_2$, H$_2$O) from a (dayside) magma ocean could then lead to a large outgassed secondary atmosphere \citep[e.g.,][]{Bower2018,Bower2019,Bower2022,Lichtenberg2021}. This atmosphere could, however, be lost because of the intense radiation from the host star \citep{Valencia2010}. Heavier outgassed species, such as Na, O$_2$ or SiO$_2$ that are outgassed from silicate melts, might be retained and form a thin silicate-vapour atmosphere \citep{Schaefer2009, Kite2016}. On the nightside, a secondary atmosphere could be built-up through volcanic outgassing from the solid mantle \citep[e.g.,][]{Kite2020,Liggins2022}. 
Observations by MOST \citep{Sulis2019}, CHEOPS \citep{Morris2021, Demory2023, MeierValdes2023}, and TESS \citep{MeierValdes2022} have shown that 55 Cancri e's occultation depth and phase curve varies with time, and the source of this variability is not yet known. Observations with JWST aim to shed light on the source of this variability and whether the atmosphere is optically thick or rather a thin-silicate vapour atmosphere \citep{PropHu2021, PropBrandeker2021}. 

55 Cancri e is one of the only super-Earths where a thermal phase curve has been observed, with the other two being lava world K2-141b \citep{Zieba2022} and the bare rock super-Earth LHS 3844b \citep{Kreidberg2019}. 
The observed brightness temperature does not necessarily correspond to the temperature at the surface of the planet. Depending on the optical depth of the atmosphere, thermal phase curve observations probe different layers of the atmosphere \citep[e.g.,][]{Parmentier2018}. Greenhouse gases such as  CO$_2$ or H$_2$ will make the atmosphere optically thick and lead to higher surface temperatures than the brightness temperature inferred from the observed outgoing long-wave radiation. We therefore use GCMs to determine the surface temperatures for different atmospheric models. The atmosphere models differ in optical thickness $\tau$ and composition (H$_2$ or N$_2$ dominated). We use the resulting longitudinal temperature profiles as a boundary condition for the surface temperature of our mantle convection models.

Thermal phase curve observations of super-Earth 55 Cancri e allow us to constrain the surface temperature but the dynamic and chemical state of its interior remains unknown. Its mass and radius measurements indicate that the planet's density is mostly consistent with either a dense and rocky planet that has a thick atmosphere, or a less dense interior with only a thin atmosphere \citep{Jindal2020,Dorn2019,Dorn2021}. 
Because of the extended pressure ranges, the convective regimes of super-Earths could be substantially different than that of Earth. \citet{Stamenkovic2012} have found that the pressure dependence of viscosity could lead to sluggish lower mantle convection and reduce the propensity for whole mantle convection. MgSiO$_3$ bridgmanite, which is the most abundant mineral in Earth's lower mantle, undergoes a phase-transition to post-Perovskite (pPv) at around $125$\,GPa \citep{Murakami2004}. For super-Earth 55 Cancri, post-Perovskite could make up most of the lower mantle. \citet{Tackley2013} investigated the effect of pPv rheology inferred from density functional theory for different sized super-Earths with Earth-like surface temperatures, and found that mantle convection is characterised by large upwellings and small time-dependent downwellings. 

Whilst the high irradiation from the host star is an important heat source to sustain a permanent magma ocean on the dayside \citep[e.g.][]{Leger2011}, the mantle also gets heated from below because the core is cooling and therefore losing heat that is left over from accretion and core formation \citep[e.g.][]{Elkins-Tanton2012, Stixrude2014}.  Additional interior heat sources include the decay of radiogenic elements, tidal heating \citep[e.g.][]{Tobie2005, Bolmont2020}, or induction heating \citep[e.g.][]{Kislyakova2017, Noack2021}. Since the amount of internal heating is difficult to constrain for super-Earths, we also investigate how the type of convection depends on the internal heating rate. The host star 55 Cancri  is an old star with an age around $8.6 \pm 1$\,Gyrs \citep{Bourrier2018}. Therefore, we also investigate the influence of a decreasing core-mantle boundary (CMB) temperature driven by core cooling.
\section{Methods} \label{sec:model}
\subsection{Planetary parameters}
55 Cancri e is a super-Earth with a radius of $1.88 \pm 0.03R_{\oplus}$ and a mass of $8.0 \pm 0.3M_{\oplus}$ \citep{Bourrier2018}. It orbits its host star in less than $18$\,hours. The time for a planet/satellite to become tidally locked scales as $\tau_{despin} \propto a^{6}$, where $a$ is the semi-major axis \citep{Peale1977,Gladman1996}. Because of its old age and ultra-short orbit, we can therefore assume that 55 Cancri e is most likely on a tidally locked orbit, leading to a strong surface temperature contrast between the dayside and the nightside. Its thermal phase curve suggests a dayside temperature of $2700 \pm 270$\,K and a nightside temperature of $1380 \pm 400$\,K \citep{Demory2016a}. This super-Earth therefore most likely harbours a magma ocean on the dayside, potentially extending towards the nightside.
We assume that the core-to-planet radius ratio is  $\approx 0.4$ \citep{Crida2018}, which leads to a core radius of $r_c = 5096$\,km and a mantle thickness of $d_{\mathrm{mantle}}=7650$\,km.
For computational reasons, we only run the models up to $4.6$\,Gyrs (age of the Sun) or less in cases where the time step has to be drastically reduced because of high velocities in the magma ocean. In most cases, a few billion years of runtime is sufficient for the model to reach a statistical steady state in terms of the location of plumes and downwellings.

\subsection{Convection Models}
We model the mantle convection of super-Earth 55 Cancri e using the code StagYY \citep{Tackley2008} in a two-dimensional (2D) spherical annulus \citep{Hernlund2008} under the infinite Prandtl number approximation (i.e. inertial forces are neglected). Compressibility is assumed by employing the truncated anelastic liquid approximation (TALA) where a reference state density profile is assumed that varies with depth. Material properties, such as thermal expansivity $\alpha$ and thermal conductivity $k$, also depend on pressure using the same parameters as in \citet{Tackley2013}. 
All models have a resolution of $256$ cells in the angular direction and $128$ cells in the radial direction, corresponding to a mean spacing of around $60$\,km in the radial direction. 
The initial temperature field has a potential temperature of $2750$\,K and a thermal boundary layer at the CMB of thickness $d_{\mathrm{TBL}}=160$\,km. 
We run models with $3$ different internal heating modes: 
\begin{itemize}
\item no internal heating: $H_0 = 0$
\item Earth-like internal heating at present-day: $H_{\mathrm{E}} =5.2 \cdot 10^{-12}$\,W/kg
\item high internal heating: $H_{\mathrm{h}} = 1.4 \cdot 10^{-11}$\,W/kg
\end{itemize}
The Earth-like internal heating rate is estimated from Earth's present-day radiogenic heating power $\approx 20$\,TW \citep{Sammon2022}. For the high internal heating case, we use an upper estimate that is appropriate for the early phase of a planet's evolution.  
For some models, we also include a simple core cooling model in which the temperature at the CMB decreases with time according to the heat flux at the CMB and assuming the heat capacity of the core $C_{p\mathrm{,core}}=750$\,J/kg\,K  \citep[][]{Labrosse2014,Gubbins2003}.

\subsection{Density and thermal expansivity}
A third order Birch-Murnaghan equation of state is used to relate density to pressure. We use a $2$-component system for the solid phase consisting of $60\%$ olivine (including 3 solid-solid phase transitions) and $40\%$ pyroxene-garnet (including 4 solid-solid phase transitions) and a $1$-component system for the melt phase. Table \ref{tab:dens_pars} shows the parameters used for the density profiles. If the melt fraction $\phi$ is between $0$ and $1$, the average density is computed using volume-additivity. Similarly, thermal expansivity is averaged volumetrically assuming the Reuss approximation \citep{Stacey1998}.

\begin{table}[h!]
\caption{Birch-Murnaghan parameters for reference density profile. }
\label{tab:dens_pars}
\centering
\begin{tabular}{cccc}
\hline\hline
Mineralogy & $K_0$ (GPa) & $K^\prime$ &$\rho_{s}$ (kg/m$^3$) \\
\hline
upper mantle    &   163 & 4.0 &  3240  \\ 
transition zone &   85  & 4.0 &  3226  \\ 
bridgmanite     &   210 & 3.9 &  3870  \\ 
post-perovskite &   210 & 3.9 &  3906  \\ 
melt            &   30    & 6.0 & 2750 \\
\hline
\end{tabular}
\tablefoot{$K_{0}$ is the bulk modulus at pressure $P=0$, $K_{0}^{\prime}$ the pressure derivative of the bulk modulus at $P=0$, and $\rho_s$ is the surface density.}

\end{table}
The thermal expansivity is calculated using 
\begin{equation} 
\alpha = \frac{\rho \gamma C_p}{K}\,,
\end{equation}
where $\gamma$ is the Grüneisen parameter, $C_p=1200$\,J/kg\,K is the heat capacity, $\rho$ the density, and $K$ the bulk modulus.

\subsection{Rheology}
For Earth, diffusion creep is the dominant deformation mechanism at higher pressures while dislocation creep is the main deformation mechanism of the upper mantle \citep{Karato1993}. In this study, we neglect dislocation creep as the majority of mantle material in 55 Cancri e is at high pressure, with the core-mantle boundary reaching over $1200$\,GPa. We employ an Arrhenius viscosity law for diffusion creep
\begin{equation}
\eta(P,T) = \eta_0\exp{\left(\frac{E_a+PV_a(P)}{RT}-\frac{E_a}{RT_0}\right)}\,, 
\end{equation}
where $P$ is pressure, $T$ is temperature, $T_0$ is the temperature at $P=0$, $E_a$ is the activation energy, $V_a(P)$ the activation volume and $R=8.31445$\,J/K\,mol the universal gas constant, and $\eta_0$ is a reference viscosity. The activation volume depends on pressure:
\begin{equation}
V_a(P) = V_0\exp{\left(-\frac{P}{P_{decay}}\right)}\,,
\end{equation}
where $P_{\mathrm{decay}}$ is the decay pressure controlling the pressure dependence of the activation volume.
The post-perovskite lower bound parameters from \citet{Tackley2013} are used: $E_a=162$\,kJ/mol, $V_0=1.40$\,cm$^3$/mol and $p_{\mathrm{decay}}=1610$\,GPa.
We use a minimum and maximum cutoff for the viscosity of $\eta_{\mathrm{min}}=10^{18}$\,Pa\,s and $\eta_{\mathrm{min}}=10^{28}$\,Pa\,s to facilitate numerical solution.
The outermost, rigid layer of a planet (lithosphere) is prone to failure if differential stresses are sufficiently large. We model this through a plastic yielding criteria. At low pressure, the strength of the lithosphere is related to its fracture strength or frictional sliding of faults \citep[Byerlee's law,][]{Byerlee1978}). At higher pressure, the strength is related to ductile failure caused by dislocation motion of the lattice \citep{Kohlstedt1995}. Both the brittle and ductile components are encapsulated within a pressure-dependent yield stress
\begin{equation}
\sigma_{\mathrm{y}} = \mathrm{min}(c + c_fP,\sigma_{\mathrm{duct}}+\sigma^{\prime}_{\mathrm{duct}}P)\,,
\end{equation}
where $c$ is the cohesive strength, $c_f$ is the friction coefficient, $\sigma_\mathrm{duct}$ the ductile yield stress at $P=0$, and $\sigma^{\prime}_{\mathrm{duct}}$ is the ductile yield stress gradient. For this study, we use $c=1$\,MPa, $c_f=0.1$, $\sigma_\mathrm{duct}=100$\,MPa, and $\sigma^{\prime}_{\mathrm{duct}}=0.01$.
If the stress exceeds the yield stress $\sigma_{y}$, the viscosity gets reduced to an effective viscosity given by  
\begin{equation} \label{chap4:eta_eff}
\eta_{\text{eff}} = \frac{\sigma_{\text{y}}}{2 \dot{\epsilon_{\text{II}}}} \quad \mathrm{if} \hspace{0.5em} 2 \eta \epsilon_{\text{II}} > \sigma_y\,,
\end{equation}
where $\dot{\epsilon_{\mathrm{II}}}$ is the second invariant of the strain rate tensor.

\subsection{GCM Simulations}
We determine the range of plausible radiative surface temperatures $T_\mathrm{{rad}}$ using simulations from the general circulation model (GCM) Exo-FMS \citep{Hammond2017,Hammond2021}. We re-ran a subset of the simulations of 55 Cancri e in \citet{Hammond2017}, choosing simulations that spanned the range of day-side radiative surface temperatures and day-night contrasts. The simulations used an updated version of the Exo-FMS GCM \citep{Hammond2021}, with a cubed-sphere grid configuration instead of the latitude-longitude grid used in \citet{Hammond2017}. The GCM setup was otherwise the same as in \citet{Hammond2017}. We used semi-grey radiative transfer where the shortwave (stellar radiation) and longwave (atmospheric emission) opacities are separately set to be constant. We used dry convective adjustment, where any unstable atmospheric temperature profiles are instantaneously adjusted to stability on the dry adiabat. The simulations were run to equilibrium as described in \citet{Hammond2017} and then time-and latitudinal-averaged radiative surface temperatures were measured.

\begin{figure}
    \centering
    \includegraphics[width=0.45\textwidth]{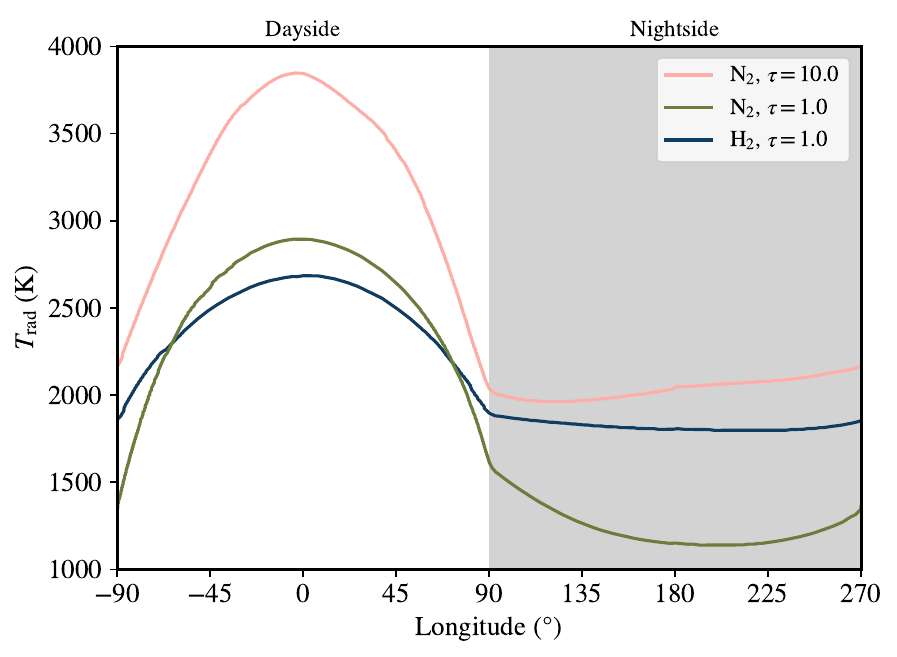}
    \caption{Radiative temperatures $T_{\mathrm{rad}}$ that are derived from general circulation models. $T_{\mathrm{rad}}$ are used to determine the surface temperature of the planet using Equation~\ref{eq:radiative_balance}. H$_2$ indicates a low molecular weight atmosphere dominated by hydrogen, and N$_2$ indicates a high molecular weight atmosphere dominated by nitrogen. $\tau$ is the optical depth of the atmosphere.}
    \label{fig:gcm_models}
\end{figure}

To explore the range of potential radiative surface temperatures in these simulations, we varied the atmospheric longwave optical depth $\tau$, and the atmospheric mean molecular weight $\mu$ (which we label by the atmospheric composition). Figure~\ref{fig:gcm_models} shows the radiative surface temperature profiles inferred from each of the GCM simulations, labelled by their molecular weight (either dominated by hydrogen H$_2$ or N$_2$) and longwave optical depth $\tau$. As expected, atmospheres with higher longwave optical depths have higher mean temperatures. \citet{Hammond2017} showed how atmospheres with lower mean molecular weights have more heat redistribution to their nightside, resulting in less temperature variation, which we see again here. Varying these two parameters has therefore shown the plausible range of surface temperatures that a thick atmosphere could produce.

\subsection{Boundary conditions}
The surface boundary condition is free-slip and radiative. The surface temperature is determined by radiative equilibrium:  
\begin{equation} \label{eq:radiative_balance}
\sigma T_s^4 = F_{\mathrm{top}} + \sigma T_{\mathrm{rad}}^{4}\,,
\end{equation}
where $\sigma = 5.670 \cdot 10^{-8}$\,Wm$^{-2}$K$^{-4}$ is the Stefan-Boltzmann constant, $F_{\mathrm{top}}$ is the surface heat flux from the interior and $T_s$ is the surface temperature.
$T_{\mathrm{rad}}$ is defined as the temperature of the planet when it is in stellar equilibrium with the irradiation from the star taking into account the effects of a potential atmosphere.

If large portions of the surface are molten, $F_{\mathrm{top}}$ can reach several hundred W/m$^2$ and the surface temperature is significantly different from the radiative equilibrium temperature. If the surface is solid, $F_{\mathrm{top}}$ is small enough such that $T_s \approx T_{\mathrm{rad}}$.
The initial CMB temperature for all models is $T_{\mathrm{CMB,init}}=9500$\,K. There is still a large uncertainty on the temperature at the CMB of super-Earths. Here, we use an estimate for the melting temperature of MgSiO$_3$ at the CMB \citep{Fei2021, Stixrude2014}. We use a lower-bound value which avoids the formation of a basal magma ocean. A basal magma ocean, whilst interesting to investigate, would add additional complexity and be deserving of its own study. Hence, we focus on the scenario of a solid mantle at the interface with the core.

\subsection{Melting}
The melt fraction of each cell is calculated as
\begin{equation}
\phi = \frac{T-T_{\mathrm{sol}}}{T_{\mathrm{liq}}-T_{\mathrm{sol}}}\,.
\end{equation}
The solidus temperature $T_{\mathrm{sol}}$ is given by \citet{Herzberg2000} for the upper mantle, \citet{Zerr1998} for the lower mantle, and \citet{Stixrude2014} for the post-perovskite phase. 
The liquidus is given by a compromise between \citet{Zerr1998}, \citet{Stixrude2009}, and \citet{Andrault2011}. The resulting solidus and liquidus curves are shown in Figure~\ref{fig:solidus_liquidus}A.

If melt is present, we parametrise the heat flux $\vec{J}_\mathrm{q}$ transported by vigorous convection in the magma ocean by assuming a very high effective thermal conductivity $k_\mathrm{h}$ (eddy diffusion) \citep{Abe1997}. $k_\mathrm{h}$ varies as a function of the melt fraction $\phi$:
\begin{equation} 
k_\mathrm{h} = \exp{\left(\frac{\ln{(k_\mathrm{{h,max}}})}{2}\left(1+\tanh{\frac{\phi - \phi_c}{\Delta\phi}}\right)\right)}-1\,,
\end{equation}
where $\phi_\mathrm{c} \approx 0.35$ is the rheological threshold (critical melt fraction) above which the mixture of solid and melt will behave rheologically as a low-viscosity fluid  \citep{Abe1997}. $\Delta\phi=0.05$ is the width of this transition between solid- and liquid-like rheological behaviour. 
For the maximum effective thermal conductivity for melt, we use $k_{\mathrm{h,max}} = 10^7$\,Wm$^{-1}$K$^{-1}$, which assumes the magma ocean is vigorously convecting. 
However, this depends on whether the temperature gradient of the magma ocean is super-adiabatic ($\frac{dT}{dP} > \frac{dT_{a}}{dP}$) or not. If the magma ocean is sub-adiabatic, it will not be convecting and therefore heat transport (especially in the vertical direction) is not necessarily enhanced. In this study, we investigate two cases: 

\begin{itemize}
	\item `adiabat-mode': The vertical heat flux $J_\mathrm{q,r}$ is enhanced only if $\frac{dT}{dP} > \frac{dT_{a}}{dP}$.
	\item `melt-mode': The vertical heat flux $J_\mathrm{q}$ is always enhanced if melt is present ($\phi > 0$).
\end{itemize}
For both modes, horizontal heat transport is enhanced using a very high effective thermal conductivity. 
`Adiabat-mode' therefore corresponds to a magma ocean that is only very efficiently transporting heat in the vertical direction if the temperature gradient is super-adiabatic. On the other hand, `melt-mode' assumes that heat is still efficiently transported even if the magma ocean is sub-adiabatic. In Section~\ref{sec:discussion_thickness}, we discuss under which circumstances this might be the case.
 We include latent heating effects ($0 < \phi < 1$) by using an effective heat capacity $C_p^{\prime}$ and an effective thermal expansion parameter $\alpha^{\prime}$ in the energy equation \citep[e.g.][]{Solomatov2007}:
\begin{equation}
C_p^{\prime} = C_p + \frac{\Delta L}{T_{\mathrm{liq}}-T_{\mathrm{sol}}}\,,
\end{equation}
\begin{equation}
\alpha^{\prime} = \alpha + \frac{\Delta \rho }{\bar{\rho} (T_{\mathrm{liq}}-T_{\mathrm{sol}})}\,,
\end{equation}
where $\Delta L$ is the latent heat of melting/freezing, $\Delta \rho$ is the density difference between the solid and melt phase and $\bar{\rho}$ is the volumetrically averaged density. To model latent heating/cooling self-consistently, we fix the latent heat of melting/freezing and the density difference between the melt and solid phase is determined using: 
\begin{equation}
\Delta \rho = \frac{\Delta L \bar{\rho}^2}{\Gamma T}
\end{equation}
Where $\Gamma$ is the Clapeyron slope of the solid-melt phase transition given by

\begin{equation} 
\Gamma = \frac{\rho g}{\phi \frac{dT_{\mathrm{liq}}(z)}{dz}+(1-\phi)\frac{dT_{\mathrm{sol}}(z)}{dz}}\,.
\end{equation}

\section{Results}

\subsection{Reference model} \label{section:ref_model}
As a reference model, we use a substellar point temperature of $T_{\mathrm{day}} = 2800$\,K and nightside temperature of $T_{\mathrm{night}}=1150$\,K. The vertical heat flux in the magma ocean is only enhanced if the temperature gradient is super-adiabatic (`adiabat-mode'). The potential mantle temperature is $2800$\,K and the mantle is basally heated (no internal heating) and the temperature of the CMB is constant: $T_{\mathrm{CMB}}=9500$\,K.

\begin{figure}
    \centering
    \includegraphics[width=0.45\textwidth]{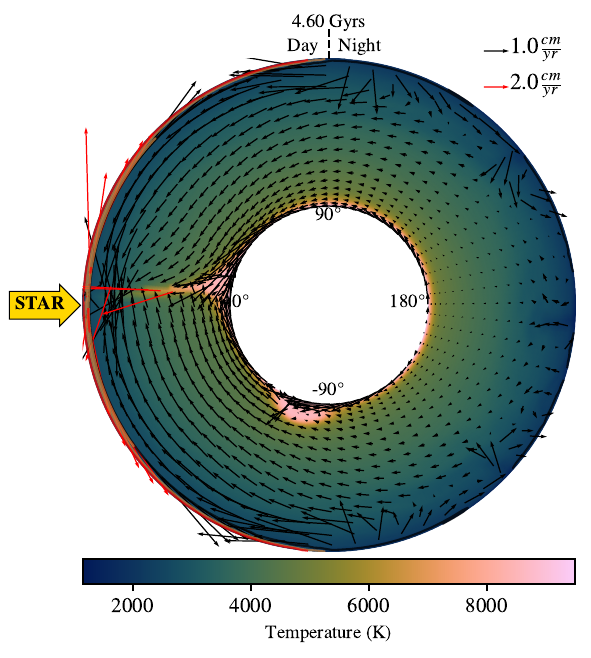}
    \caption{Mantle temperature field and velocity field of super-Earth 55 Cancri e for the reference model. High velocities (larger than three times the mean mantle velocity) are plotted with a red arrow. The magma ocean is indicated in red (fully molten) and orange (partially molten).}
    \label{fig:cnc_1}
\end{figure}

Figure~\ref{fig:cnc_1} shows the temperature field at $4.6$\,Gyrs for this model. A prominent plume has formed on the dayside of the planet and there is a flow of solid material from the nightside towards the dayside (indicated by the vector field). The return flow from the dayside towards the nightside is accommodated by the magma ocean and the uppermost solid part of the mantle. Velocities in the (near) magma ocean region are much higher than in the  solid mantle. We therefore use a different scale for the high velocities (red arrows). That is, velocities that are larger than $3$ times the mean mantle velocity: 
\begin{equation}
\lVert v(r,\phi) \rVert > 3 \left\langle\lVert v(r,\phi)\rVert\right\rangle_\mathrm{mantle}\,.
\end{equation}

\begin{figure*}
    \centering
    \includegraphics[width=0.8\textwidth]{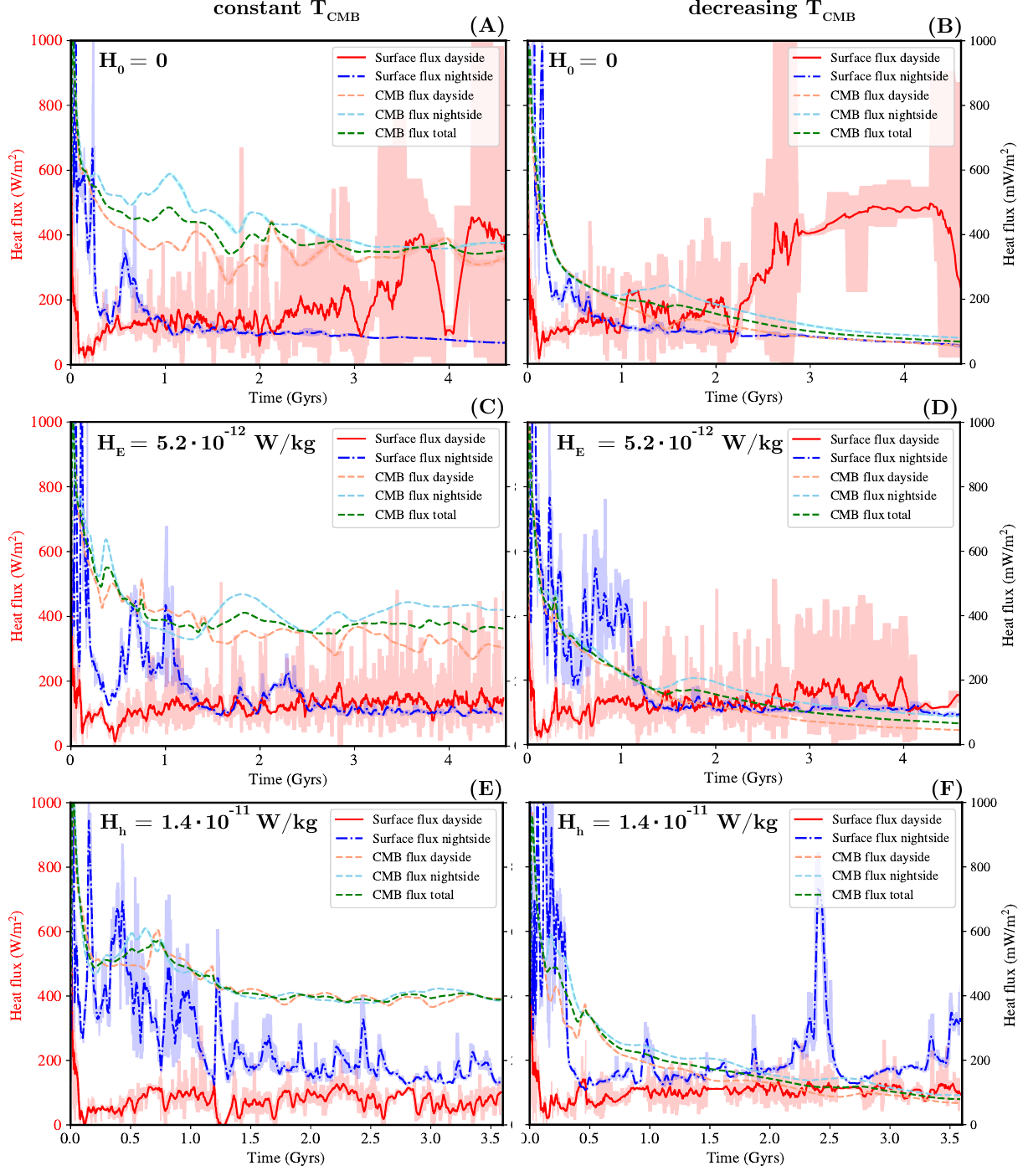}
    \caption{Surface and CMB heat flux for the 55 Cancri e models with $T_{\mathrm{day}}=2800$\,K and $T_{\mathrm{night\mathrm}}=1150$\,K. The nightside heat flux (shown in blue dot-dashed) and the CMB heat flux (dashed lines) have units of mW/m$^{2}$ (right axis). The red solid line shows the dayside surface heat flux in W/m$^{2}$ (left axis). The shaded regions on either side of the lines represent the minimum and maximum values within a moving average window. $H_{0}$, $H_{\mathrm{E}}$, and $H_{\mathrm{h}}$ indicate no, Earth-like, and high internal heating respectively. If the CMB temperature is constant, the heat flux from the CMB stabilises after roughly $1$\,Gyr. There is more variability in the surface heat flux for both the dayside and nightside heat flux. For all cases, the dayside and nightsides surface have thermal evolutions that are distinct from each other and the dayside heat flux is around $3$ orders of magnitude higher than the heat flux coming from the (partially molten) nightside.}
    \label{fig:cnc_ref_heatflux}
\end{figure*}

At the beginning of the model, the magma ocean has a depth of around $3500$\,km on the dayside (fully molten) and $2700$\,km on the nightside (partially molten). At $4.6$\,Gyrs, the magma ocean has a thickness of $440$\,km on the dayside (fully molten) and the nightside is mostly solid.
Figure~\ref{fig:cnc_ref_heatflux}A shows the corresponding surface and CMB heat flux. The dayside is fully molten and therefore has a much greater heat flux than the nightside. The heat flux at the CMB is $\approx 375$\,mW/m$^2$. 

\subsection{Role of internal heating and core cooling}

\begin{figure*}
    \centering
    \includegraphics[width=0.8\textwidth]{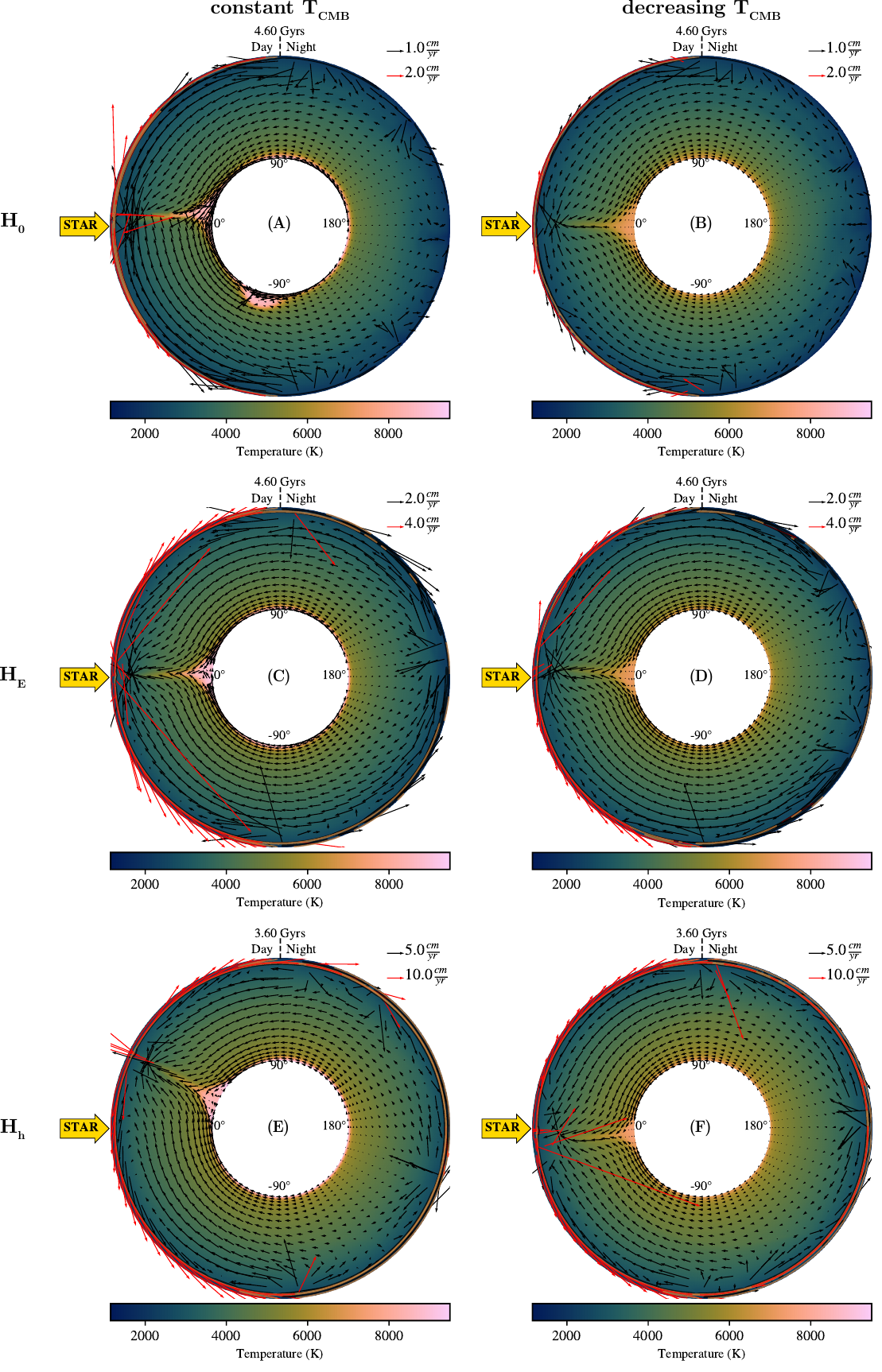}
    \caption{Snapshots of mantle temperature and velocity fields for the 55 Cancri e models with $T_{\mathrm{day}}=2800$\,K and $T_{\mathrm{night}}=1150$\,K. High velocities (larger than three times the mean mantle velocity) are indicated by red arrows. The magma ocean is indicated in red (fully molten) and orange (partially molten). $H_{0}$, $H_{\mathrm{E}}$, and $H_{\mathrm{h}}$ indicate no, Earth-like, and high internal heating respectively. The left column shows the models with a constant CMB temperature. The right column shows the models where the CMB temperature decreases due to core cooling. For all models, a magma ocean forms on the dayside and the nightside is partially molten.  The most prominent feature is the strong upwelling that is present on the dayside.} 
    \label{fig:cnc_ref_Tmodels}
\end{figure*}

In this section, the parameters are the same as for the reference model, but we now apply different internal heating rates, and the models have either a constant CMB temperature or decreasing CMB temperature due to core cooling.
The internal heating rate is $H_0=0$ (no internal heating),  $H_{\mathrm{E}}=5.2\cdot10^{-12}$\,W/kg (present-day Earth-like internal heating), and $H_{\mathrm{h}}=1.4\cdot10^{-11}$\,W/kg (high internal heating). Models with high internal heating have only been run up to $3.6$\,Gyrs, because the higher velocities require smaller time steps.
Figure~\ref{fig:cnc_ref_Tmodels} shows the temperature and velocity fields of the mantle after $4.6$\,Gyrs (or $3.6$\,Gyrs for the model with high internal heating) and Figure~\ref{fig:cnc_ref_etamodels} shows the corresponding viscosity fields. In all cases, a prominent upwelling forms on the dayside and there is a flow of solid material from the nightside towards the dayside. 
Small downwellings form in the bridgmanite layer for the model without internal heating, but they are not strong enough to penetrate through the bridgmanite-pPv phase transition where they become more diffuse. 
Solid material on the nightside moves more slowly than on the dayside because of the temperature-dependent viscosity. The return flow from the dayside towards the nightside is accommodated in the near-surface layer, which is molten at the surface on the dayside for all models and partially molten on the nightside. The partially molten magma-ocean on that side is, however, very thin, with a thickness of less than $60$\,km.

\begin{figure}
    \centering
    \includegraphics[width=0.45\textwidth]{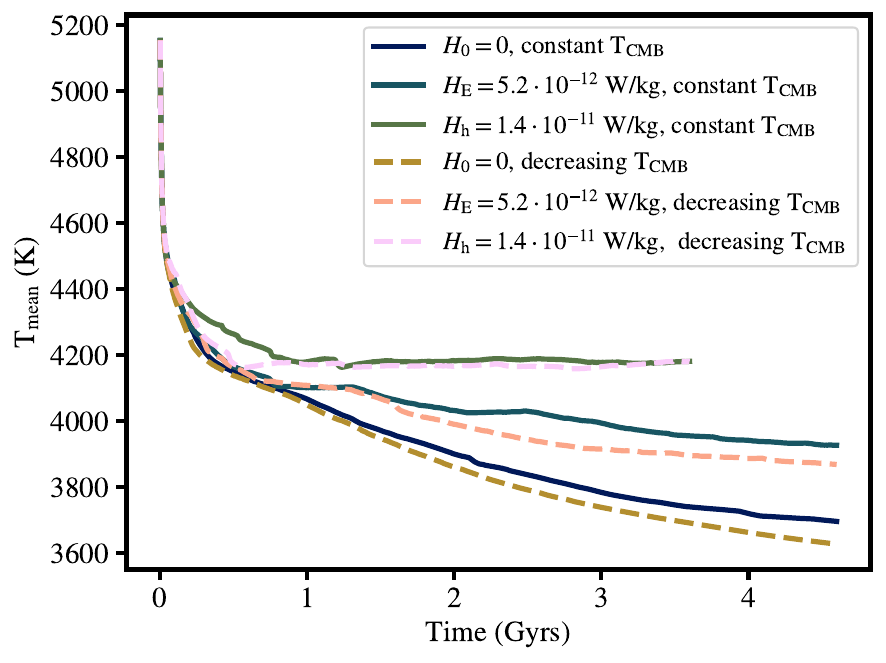}
    \caption{Mean mantle temperature for different models of 55 Cancri e with different rates of internal heating and constant or decreasing CMB temperature. The dashed lines indicate models that include decreasing CMB temperature due to core cooling. Solid lines indicate models with constant CMB temperature. The mantle temperature stabilises after around $1$ and $4$\,Gyrs for the models with high and Earth-like internal heating respectively. Without internal heating, the mantle temperature will continue to decrease beyond $4.6$\,Gyrs.}
    \label{fig:Tmean_cnc_ref}
\end{figure}

Figure~\ref{fig:Tmean_cnc_ref} shows the mean mantle temperature for the different reference model cases (i.e., $T_{\mathrm{day}}=2800$\,K and $T_{\mathrm{night}}=1150$\,K). The models with high internal heating stabilise around $1$\,Gyr with a mean mantle temperature of $4200$\,K. Comparing the models with either constant or decreasing CMB temperature, the mean mantle temperature varies by less than $100$\,K.
With an Earth-like internal heating rate, both models reach steady state after around $4\,$Gyrs and the mantle temperature is around $50$\,K cooler if the temperature at the CMB is decreased because of core cooling. 
The model without internal heating cools the most and will continue to cool beyond $4.6$\,Gyrs. The mean mantle temperature at that time is  $3630$\,K if the CMB temperature decreases because of core cooling, and $3700$\,K if the CMB temperature is constant. The results of our models show that including the effect of core cooling on the CMB temperature does not change the mean mantle temperature or the mantle dynamics significantly.   
With a high internal heating rate, the mantle temperature remains constant after $\approx1.2$\,Gyrs at $4200$\,K independent of whether the temperature at the CMB is constant or decreasing.

Figure~\ref{fig:cnc_ref_heatflux} shows the surface and CMB heat flux for the different models.  
For the models without core cooling, the CMB heat flux after $4.6$ Gyrs (or $3.6$\,Gyrs for the case with high internal heating) is $\approx 355$\,mW/m$^{2}$, $\approx 363$\,mW/m$^{2}$, and $\approx 389$\,mW/m$^{2}$ for no, Earth-like, and high internal heating, respectively. 
When the CMB temperature is decreasing, the CMB heat flux will also steadily decrease over time. With no internal heating the flux at $4.6$\,Gyrs is $\approx 68$\,mW/m$^{2}$, with Earth-like internal heating it is $\approx 62$\,mW/m$^{2}$ and for high internal heating it is slightly higher at $\approx 78$\,mW/m$^{2}$ (at $3.6$\,Gyrs). 
The dayside surface flux is roughly $3$ orders of magnitude larger than the nightside surface flux for all model cases, which is why it is shown using a separate axis in W/m$^{2}$. This is because the dayside harbours a fully molten magma ocean, whereas the nightside is only partially molten or solid. 

In summary, we observe for all models that a magma ocean forms on the dayside of the planet and the nightside is partially molten. The most prominent feature is the strong upwelling that is present on the dayside for all models (i.e. independent of the amount of internal heating and whether the CMB temperature is constant or not). If upwellings form around the CMB, they tend to move towards the dayside and merge with the upwelling that is already present. Solid mantle material moves from the nightside surface into the mantle and then towards the dayside. The return flow of material from the dayside towards the nightside is accommodated in the magma ocean and uppermost layer of the solid mantle where the viscosity is the lowest.

\subsection{Role of dayside and nightside temperature} \label{sect:surf_t}
\begin{figure*}[h!]
    \centering
    \includegraphics[width=0.8\textwidth]{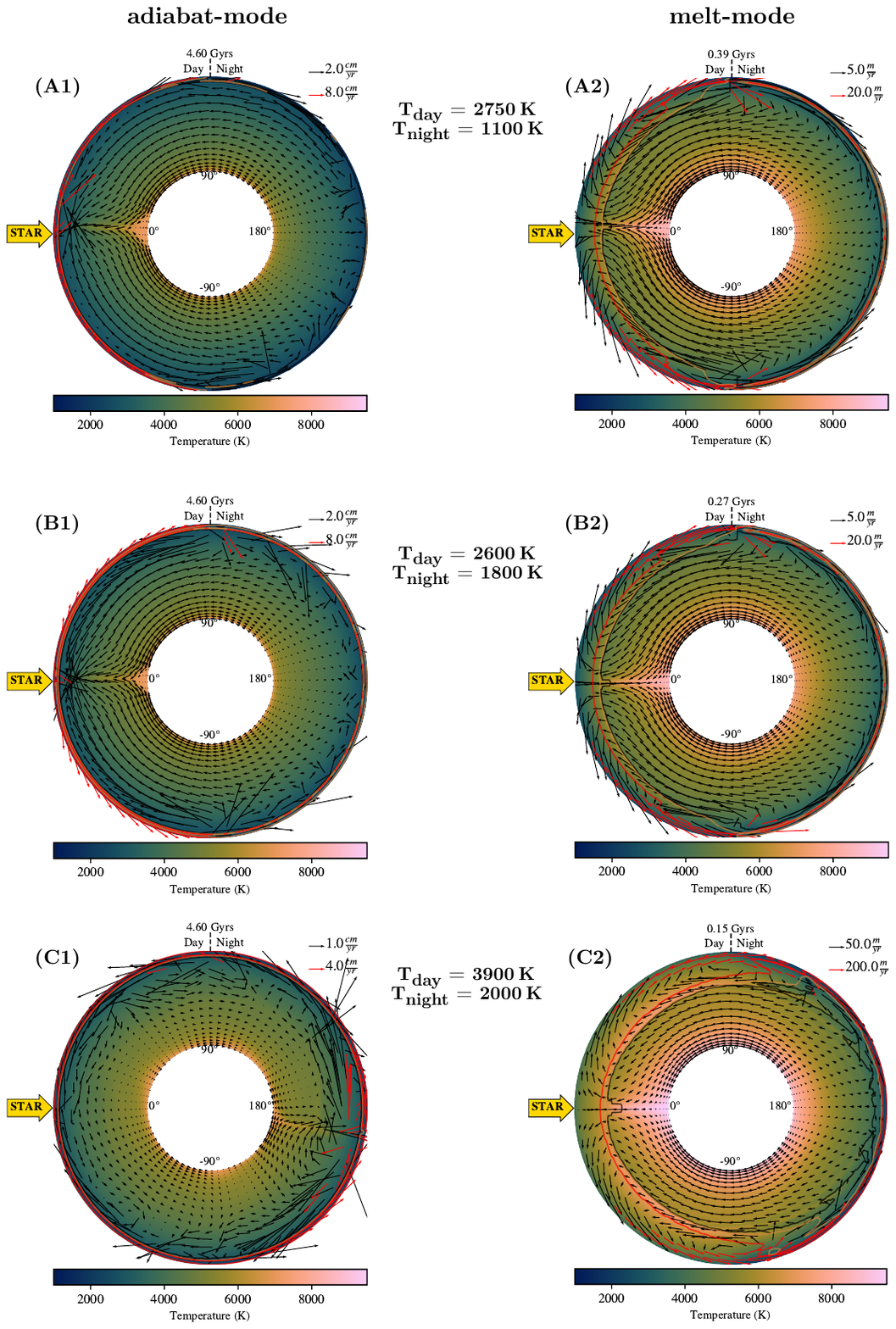}
    \caption{Snapshots of mantle temperature for the `adiabat-mode' (left column) and `melt-mode' (right column) for different models of 55 Cancri e with different surface temperature contrasts. High velocities (larger than three times the mean mantle velocity) are indicated by red arrows. The magma ocean is indicated in red (fully molten) and orange (partially molten). If the nightside is not or only partially molten, an upwelling forms that is preferentially on the dayside (A,B). If both the dayside and nightside are fully molten, the upwelling does not have a preferred location anymore and can also be stable on the nightside (C).}
    \label{fig:cnc_tsurf_models}
\end{figure*}

\begin{figure*}[h!]
    \centering
    \includegraphics[width=0.8\textwidth]{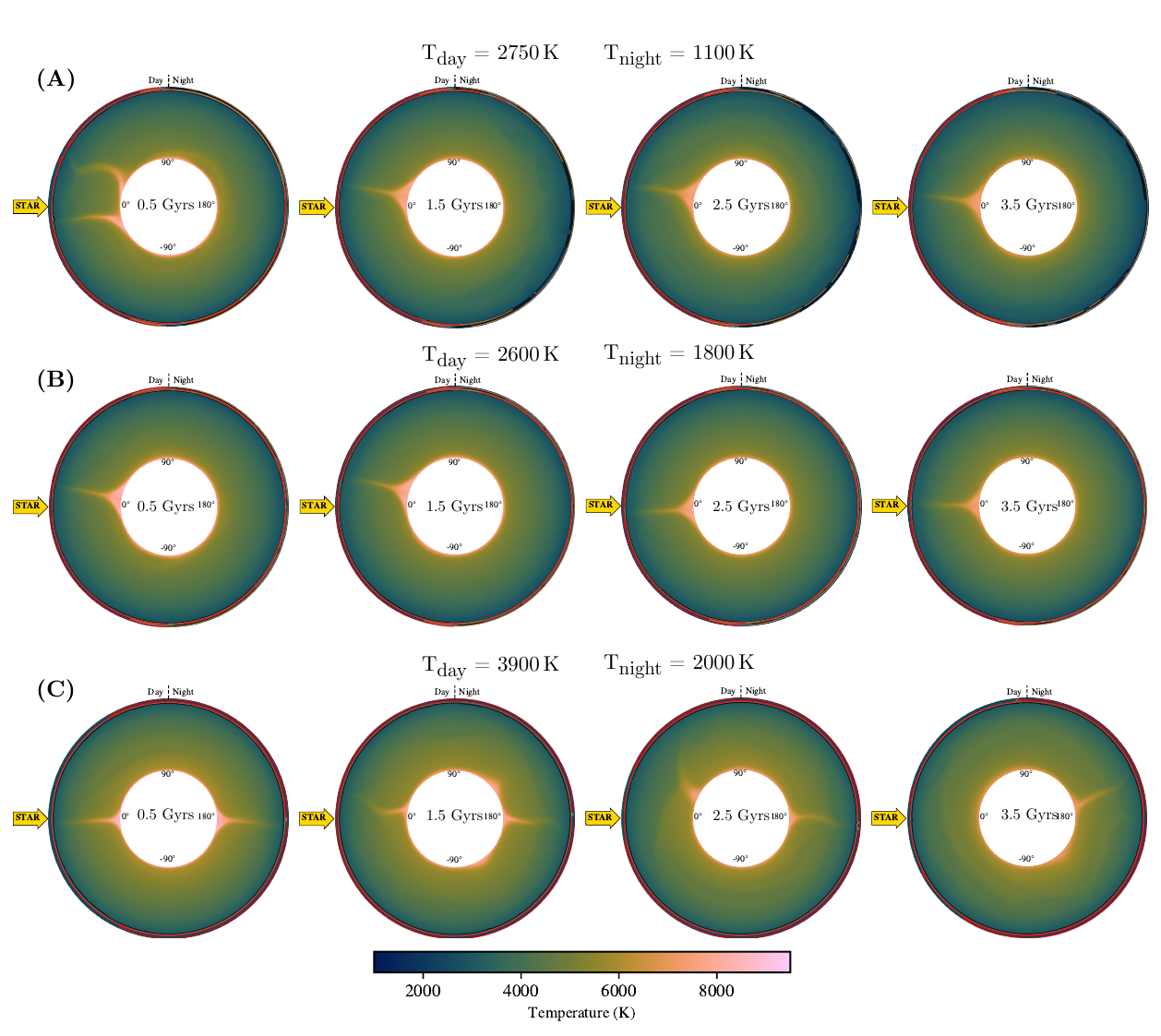}
    \caption{Snapshots of mantle temperature of 55 Cancri e at different time steps for the models with different surface temperature contrasts. The magma ocean is indicated in red (fully molten) and orange (partially molten). If the nightside temperature remains low enough for the surface not to be fully molten, a strong upwelling is preferentially on the dayside (A,B). If the nightside is fully molten, upwellings form at different locations around the CMB and some will also move towards the nightside (C).}
    \label{fig:cnc_tsurf_evo}
\end{figure*}

\begin{figure}
    \centering
    \includegraphics[width=0.45\textwidth]{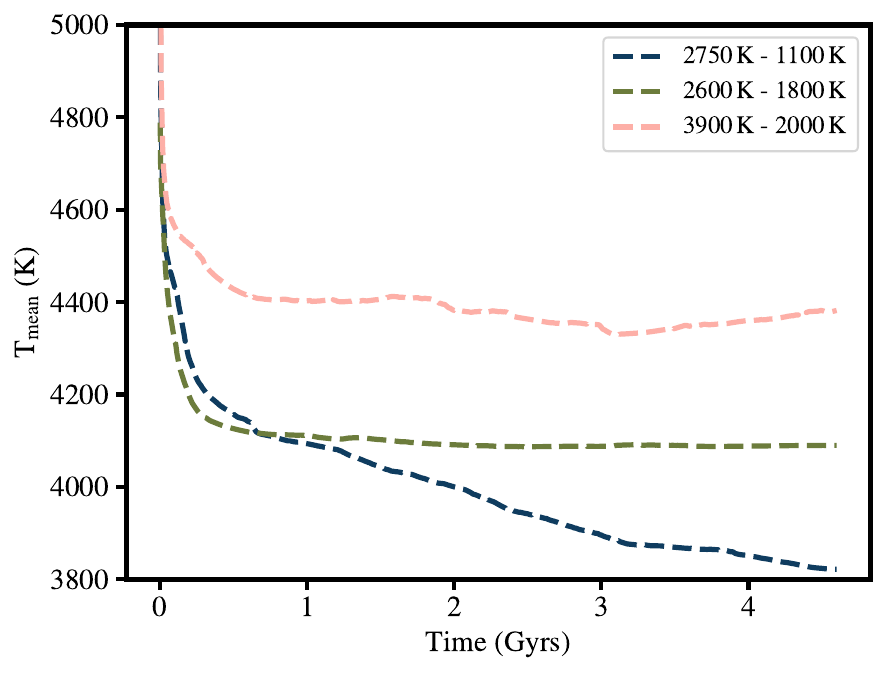}
    \caption{Mean temperature of the mantle of 55 Cancri e for models with different radiative temperatures $T_{\mathrm{rad}}$ that are derived from general circulation models. The models all have Earth-like internal heating and decreasing CMB temperature. The models with higher temperatures reach steady state after around $1$Gyr, whereas the mantle temperature for the model with the lowest nightside temperature continues to decrease beyond $4.6$\,Gyrs.}
    \label{fig:cnc_tsurf_tmean}
\end{figure}

Here, we show the results for the interior dynamics of 55 Cancri e for different surface temperature contrasts. The different radiative temperature profiles are taken from the general circulation models and are shown in Figure~\ref{fig:gcm_models}. All models feature Earth-like internal heating ($H_{\mathrm{E}}=5.2\cdot10^{-12}$\,W/kg) and the CMB temperature is decreased because of core cooling. 
Figure~\ref{fig:cnc_tsurf_models} (left column) shows snapshots of the mantle temperature at $4.6$\,Gyrs. The corresponding viscosity fields are shown in the appendix (Fig.~\ref{fig:cnc_tsurf_eta}). Figure~\ref{fig:cnc_zebra_tsurf} shows the corresponding evolutionary tracks of upwellings and downwellings. 
The model with a dayside temperature of $T_\mathrm{day}=2750$\,K and nightside temperature of $T_\mathrm{night}=1100$\,K is very similar to the reference case, but the dayside temperature is slightly cooler. A strong upwelling forms on the dayside of the planet and cold material descends into the deep mantle on the nightside (Fig.~\ref{fig:cnc_tsurf_models}: A1). 
Figure~\ref{fig:cnc_tsurf_evo}A shows snapshots of the mantle temperature at different times. A second plume forms around $0.2$\,Gyrs which then moves towards the dayside and merges with the dominant plume that is already present on that side. No more plumes form after this merging, which is completed after roughly $0.8$\,Gyrs. The plumes that are shown in the evolutionary tracks plot between around $1$ and $2$\,Gyrs (Fig.~\ref{fig:cnc_zebra_tsurf}A) correspond only to a slightly thickened thermal boundary layer. 
At $4.6$\,Gyrs, the magma ocean on the dayside has a thickness of $440$\,km and the nightside surface is mostly solid. 
The mean mantle temperature after $4.6$\,Gyrs is $3820$\,K and the mantle is still continuing to cool (Fig.~\ref{fig:cnc_tsurf_tmean}). 
    
For the model with a dayside temperature of $T_\mathrm{day}=2600$\,K and nightside temperature of $T_\mathrm{night}=1800$\,K, an upwelling also forms on the dayside and cold, solid material moves into the deep mantle on the nightside (Fig.~\ref{fig:cnc_tsurf_models}: B1).
The plume on the dayside is stable over several Gyrs (Figs.~\ref{fig:cnc_zebra_tsurf}B, \ref{fig:cnc_tsurf_evo}B). No additional plumes form once the dominant plume on the dayside has formed; rather, the lower boundary layer continuously drains through the established plume (Fig.~\ref{fig:cnc_zebra_tsurf}B).
At $4.6$\,Gyrs, the magma ocean on the dayside has a thickness of around $495$\,km and around $380$\,km on the nightside. Although the surface is only partially molten on the nightside. The mean mantle temperature after $4.6$\,Gyrs is stable at $4090$\,K. This temperature is reached after roughly $0.6$\,Gyrs (Fig.~\ref{fig:cnc_tsurf_tmean}). 

The model with a dayside temperature of $T_\mathrm{day}=3900$\,K and nightside temperature of $T_\mathrm{night}=2000$\,K has the hottest dayside temperature that we investigated and also features the strongest temperature contrast. In this case, the magma ocean covers the whole planet, which causes the temperatures inside the magma ocean to be equilibrated on a timescale that is much faster than the equilibration timescale of the underlying solid mantle. The interface between the molten and solid phase therefore acts as a constant thermal boundary condition for the solid mantle.  The plumes now form uniformly around the CMB and do not move towards the dayside anymore. At $4.6$\,Gyrs, a prominent plume is present on the nightside (Fig.~\ref{fig:cnc_tsurf_models}: C1). 
 However, this plume is not as stable as in the cases where the nightside is cool enough to be solid or partially molten with a prominent plume that forms on the dayside. Figure~\ref{fig:cnc_zebra_tsurf}C shows that the dominant plume on the dayside starts to move towards the nightside at around $2.5$\,Gyrs. 
 Figure~\ref{fig:cnc_tsurf_evo}C shows the mantle temperature at $4$ different times: At $0.5$\,Gyrs, two plumes have formed on the dayside and nightside. At around $1.5$\,Gyrs, two more plumes are forming on the nightside which then merge with the already present plume on that side. At $2.5$\,Gyrs, the plume on the dayside starts to move towards the nightside and merges with the plume on the nightside at around $3$\,Gyrs.
 The mantle temperature reaches a stable state after roughly $0.6$\,Gyrs. The mean mantle temperature after $4.6$\,Gyrs is $4380$\,K.

\begin{figure*}[h!]
    \centering
    \includegraphics[width=0.8\textwidth]{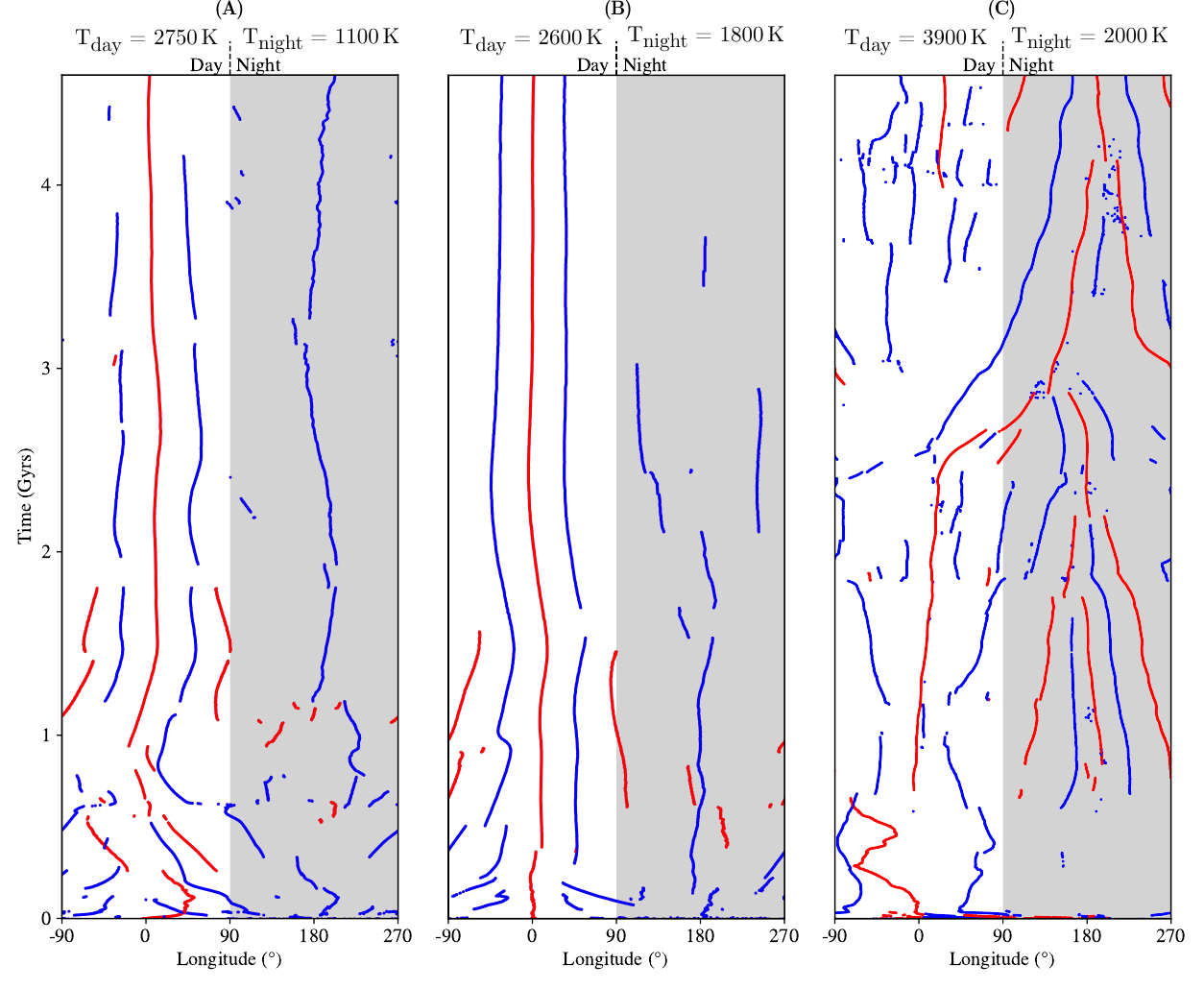}
    \caption{Evolutionary tracks showing the longitude of upwellings (red) and downwellings (blue) as a function of time for the models with different surface temperature contrasts. The nightside is indicated in grey. If the nightside temperature remains low enough for the surface not to be fully molten, a strong upwelling is preferentially on the dayside (A,B). If the nightside is fully molten, upwellings form at different locations around the CMB and some will also move towards the nightside (C).}
    \label{fig:cnc_zebra_tsurf}
\end{figure*}

\begin{figure*}[h!]
    \centering
    \includegraphics[width=0.5\linewidth]{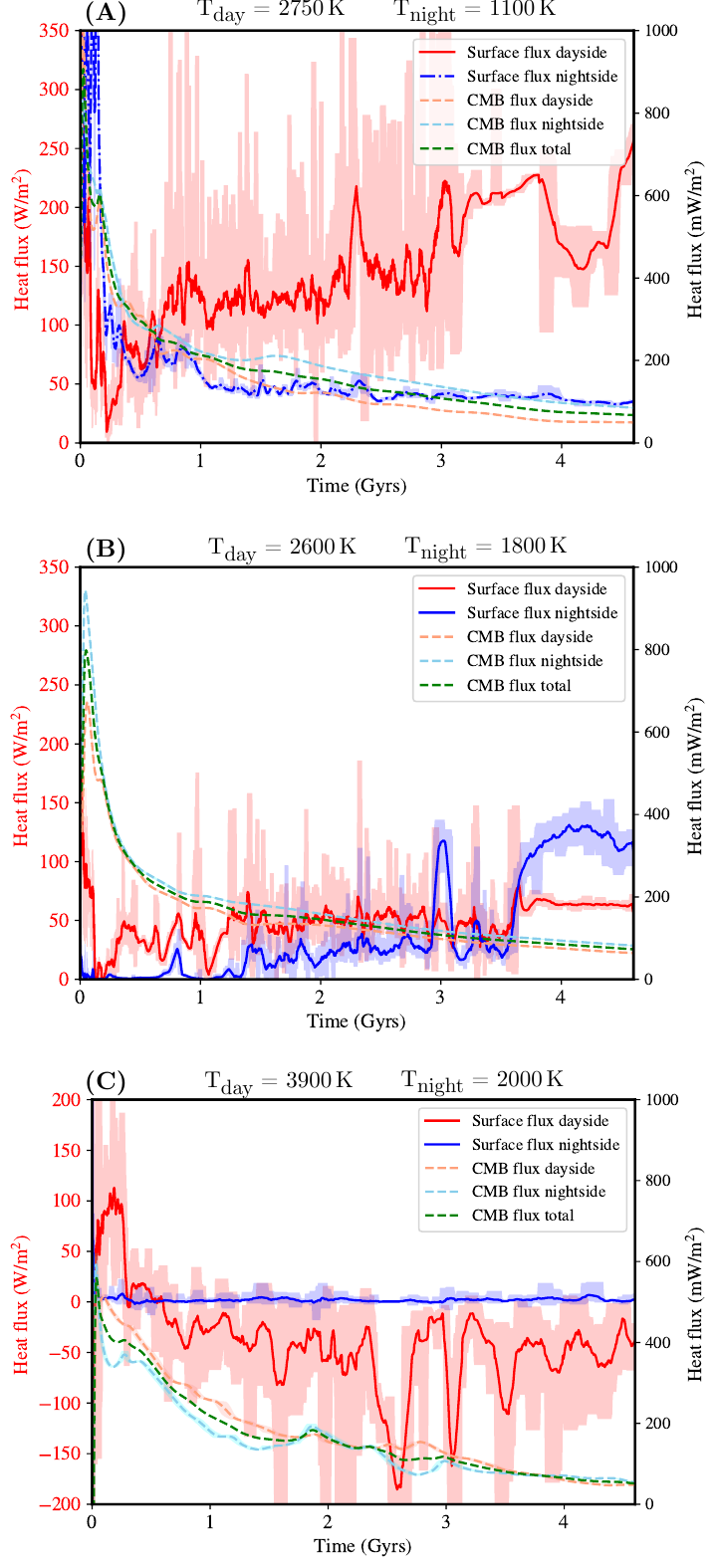}
    \caption{Surface and CMB heat flux for models with different surface temperature contrasts. The dashed lines are in mW/m$^{2}$ (right axis) and the solid lines in W/m$^{2}$ (left axis). The shaded regions on either side of the lines represent the minimum and maximum values within a moving average window. If the nightside is not fully molten, the dayside heat flux is roughly 3 orders of magnitude higher than the nightside heatflux (A). For the case with $T_\mathrm{day}=2600$\,K and $T_\mathrm{night}=1800$\,K, the surface fluxes are on the same order of magnitude (B). For $T_\mathrm{day}=3900$\,K and $T_\mathrm{night}=2000$\,K, the nightside heat flux is varying between $1$--$4$\,W/m$^2$ whereas the dayside heat flux varies from $0$ to $-200$\,W/m$^2$ (C).}
    \label{fig:cnc_tsurf_heatflux}
\end{figure*}

 Figure~\ref{fig:cnc_tsurf_heatflux} shows the surface and CMB heat flux for the different models. For the model with $T_\mathrm{day}=2750$\,K and $T_\mathrm{night}=1100$\,K, the heat flux on the dayside varies between $50$ and $200$\,W/m$^{2}$ and the nightside heat flux is steadily decreasing, reaching roughly $100$\,mW/m$^{2}$ after $4.6$\,Gyrs. The CMB heat flux after $4.6$\,Gyrs is around $68$\,mW/m$^{2}$.
 For the model with $T_\mathrm{day}=2600$\,K and $T_\mathrm{night}=1800$\,K, the dayside heat flux is around $50$\,W/m$^{2}$ and the nightside heat flux around $25$\,W/m$^{2}$ for most of the time, but increases to around $100$\,W/m$^{2}$ after $3.6$\,Gyrs. The CMB heat flux after $4.6$\,Gyrs is around $73$\,mW/m$^{2}$.
 For the model with $T_\mathrm{day}=3900$\,K and $T_\mathrm{night}=2000$\,K, the dayside heat flux varies from $0$ to $-200$\,W/m$^{2}$. A negative heat flux means that heat is flowing into the mantle. From Figure~\ref{fig:cnc_tsurf_tmean}, we can see, however, that there is no significant heating of the mantle. Heat flowing into the mantle through a near-surface temperature inversion (which depends on the temperature gradient) is not efficient compared to the energy that is radiated away from the surface (which depends on the temperature to the fourth power).  
 The nightside heat flux is around $1$--$4$\,W/m$^{2}$. The CMB heat flux after $4.6$\,Gyrs is approximately $51$\,mW/m$^{2}$.
 
In summary, we find that a large super-plume forms on the dayside of the planet if the dayside harbours a magma ocean and the nightside remains solid or is partially molten. Cold material moves into the interior on the nightside and towards the dayside. If the nightside also harbours a magma ocean, the temperature at the interface between the magma ocean and the underlying solid mantle equilibrates, so that plumes and downwellings do not have a preferred location anymore. 
 
\subsection{Efficiency of magma ocean heat transport}
As we have seen in the models using the `adiabat-mode', the magma ocean’s maximum thickness is at most $500$\,km, which corresponds to $7$\% of the whole mantle depth (Fig.~\ref{fig:cnc_tsurf_models}: left column). Because of the high stellar irradiation, the temperature profile in the magma ocean is mostly sub-adiabatic and therefore not vigorously convecting. Here, we instead investigate the case where heat transport in the magma ocean is always efficient if melt is present and how this affects the magma ocean thickness (`melt-mode').  

Figure~\ref{fig:cnc_tsurf_models} (right column) shows snapshots of the mantle temperature for the different models. The vector plot shows the velocity field and the contour plot shows the melt fraction $\phi$ ($\phi=1$ in red, $\phi=0.5$ in orange, and $\phi=0.0$ in black).
The temperatures are higher than in the case where the vertical heat flux gets only enhanced for super-adiabatic cells. This leads to lower viscosities (Fig.~\ref{fig:cnc_tsurf_eta}: A2, B2, C2) and higher velocities, which is why we did not run the models to $4.6$\,Gyrs. However, the models have reached a steady state after around $100$--$200$\,Myrs. 

\begin{figure}
    \centering
    \includegraphics[width=0.45\textwidth]{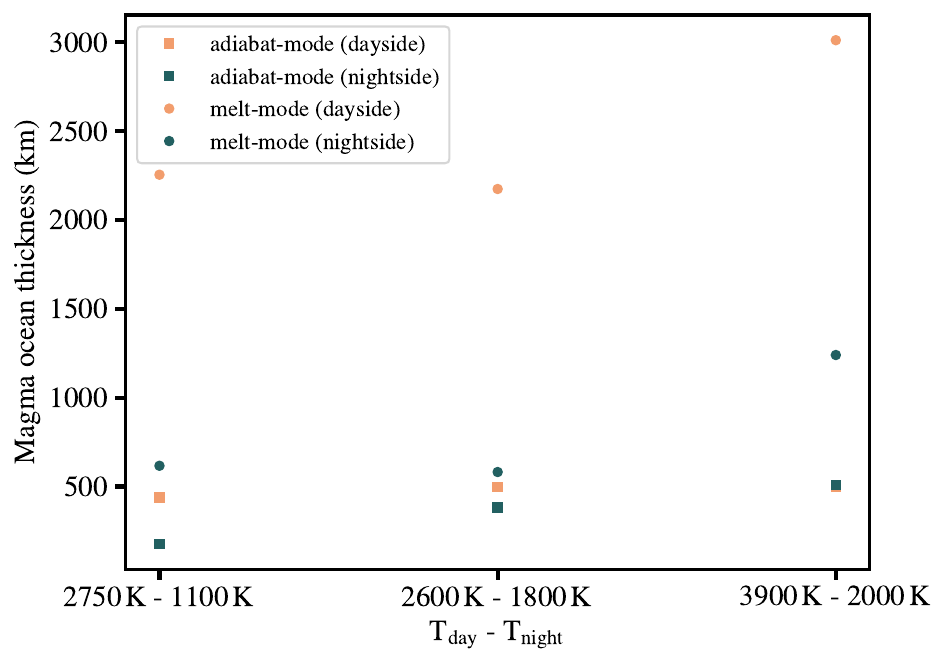}
    \caption{Magma ocean thickness for models with different surface temperature contrasts. The box markers show the thickness for models where vertical heat transport is only enhanced if the temperature gradient inside the magma ocean is super-adiabatic. The round markers show the depth for models where heat transport is always enhanced inside the magma ocean. The magma ocean thickness was determined at $4.6$\,Gyrs (or at the end of run if the model did not reach that time).}
    \label{fig:mo_depth_allmodels}
\end{figure}

If we allow the magma ocean to always transport heat very efficiently (even if it is sub-adiabatic) the magma ocean will reach depths ranging from $2000$\,km to $3000$\,km which is roughly $5$ times deeper than the thickness of the magma oceans where vertical heat transport is suppressed if it is sub-adiabatic (Fig.~\ref{fig:mo_depth_allmodels}).

\section{Discussion} \label{sec:discuss}
\subsection{Convective regime}
Our results show that a degree-1 convection pattern is established if the dayside is molten and the nightside is solid (or only partially molten). A super-plume (also called mega-plume; see Section~\ref{sect:super_plume}) preferentially forms on the dayside and cold material descends into the interior on the nightside and is then transported towards the dayside. The return flow of material from the dayside towards the nightside is accommodated in the magma ocean and the uppermost solid mantle. 
Hence, the strong surface temperature contrast provides a mechanism to stabilise deep mantle features such as plumes, which are anchored in place if a hemispheric magma ocean forms. This is is similar to how plumes are shepherded towards one hemisphere in purely rocky super-Earths with strong surface temperature contrasts, such as LHS 3844b \citep{meier2021}.
Once the super-plume has been established, the lower thermal boundary layer preferentially drains through this super-plume. This prevents the formation of additional plumes and further stabilises the degree-1 convection pattern.

For models with higher nightside temperatures, where the nightside also becomes fully molten, we find that the convection pattern is now more similar to a regime with a uniform surface temperature with plumes forming all around the core mantle boundary layer. These plumes do not have a preferred location anymore and a degree-2 convection pattern is also a stable regime (Figure~\ref{fig:cnc_tsurf_evo}C). A (super-adiabatic) magma ocean that covers the whole planet will be very efficient in redistributing heat and will equilibrate temperatures very quickly. Because the timescale for this equilibration is much faster than the equilibration timescale of the underlying solid mantle, this creates a boundary condition for the latter that is basically uniform. 
For the cases where heat transport is always efficient even if it is sub-adiabatic, we observe a degree-1 convection pattern for all cases (Fig.~\ref{fig:cnc_tsurf_models}: A2, B2, C2). The reason for this could be that the magma ocean on the dayside is almost $5$ times as deep as on the nightside, which could again provide a mechanism for the upwelling to be preferentially on the dayside. 

For lower surface temperatures without internal heating, some small downwellings form in the bridgmanite layer. However, these downwellings are not able to penetrate deep into the mantle through the bridgmanite-Ppv phase transition (Fig.~\ref{fig:cnc_ref_etamodels}). For higher surface temperatures and higher internal heating, these downwellings become more diffuse. Overall, we find that convection is rather sluggish and that convective vigour is greatly reduced by the pressure-dependence of viscosity \citep{Stamenkovic2012}.
A future study could investigate whether for smaller super-Earths these downwellings could accumulate enough buoyancy to penetrate through the bridgmanite-pPv interface.

\subsection{Super-Plume} \label{sect:super_plume}
Our models show that so-called super-plumes can form in the mantle. If the dayside harbours a magma ocean and the nightside is solid or partially molten, this super-plume will preferentially be on the dayside. The plume is rather stable located at the substellar point ($0^{\circ}$), although some of the models show that it can move up to $30^{\circ}$ east or west within approximately $1$\,Gyr (e.g. Fig.~\ref{fig:cnc_zebra_tsurf}A). 
The formation and location of the super-plume does not depend on the different efficiencies of melt transport. Therefore, the formation of a super-plume on the dayside hemisphere is a robust result.  
On Earth, volcanic hotspots, such as Hawaii or Iceland are believed to originate from deep mantle plumes \citep{Morgan1971}. Because Earth's plates move over such comparatively stationary mantle plumes, volcanoes will drift away from their plume origin, becoming inactive and forming volcanic island chains. 55 Cancri e, however, has a molten (dayside) surface, and it is unclear how the underlying mantle plume on that hemisphere would interact with the magma ocean, particularly in terms of chemical exchange. For this, our simplified approach of modelling the magma ocean by assuming enhanced heat transport through a very high eddy diffusivity becomes limited. At the interface where the solid plume is in contact with the liquid magma ocean, the heat carried by the plume will marginally heat the magma ocean, but on the other hand latent heat will be absorbed when the solids start melting. However, since we do not model in high resolution the solid-liquid interactions between the magma ocean and the plume, our models cannot specifically inform how volcanism on a magma ocean world operates. A future study could therefore focus on the local interactions between the solid plume and the magma ocean.
 
The super-plume in our model has its origins in the deep mantle and it forms because of the lower thermal boundary layer instability at the core mantle boundary. The plume therefore flushes the chemical contents of the lower hot thermal boundary layer into the magma ocean \citep{1998GeoRL..25.1999T}. If this planet has an enriched reservoir of highly volatile species (e.g., water, carbon, nitrogen, sulfur) comparable to \citep{Broadley2022,Krijt2022} or larger than \citep{Dorn2021,Lichtenberg2022a} Earth, then the super-plume would act as a funnel to transport these volatiles into the magma ocean where they would subsequently be outgassed into the atmosphere \citep[e.g.,][]{Schaefer2012,Kite2016,Gaillard2022}. Furthermore, due to the plume's proximity to the iron core, chemical products from core-mantle interactions may also be swept upwards towards the surface, enhancing chemical exchange between the core, mantle, and ultimately residual atmosphere \citep{Lichtenberg2021b,Schlichting2022}.  At around $0.9$\,TPa, MgSiO$_3 $ post-perovskite could undergo another dissociation transition \citep{Umemoto2011} which could lead to a small-scale convective layer above the CMB which promotes the formation of super-plumes \citep{Shahnas2023}.  Including this transition in our models could therefore reduce the degree of convection for the cases where the nightside is completely molten (e.g. Fig.~\ref{fig:cnc_tsurf_evo}C).

\subsection{Observations}
The thermal phase curve of 55 Cancri e shows a large offset of the hot spot $41^{\circ}$ east of the substellar point \citep{Demory2016a}. 
A recent reanalysis of the Spitzer observations has, however, also suggested that 55 Cancri e only has a negligible hot spot shift of $-12^{\circ}$ east of the substellar point \citep{Mercier2022}. 
For some of our models, (e.g. Fig.~\ref{fig:cnc_ref_Tmodels}E), we see that the super-plume is considerably shifted towards the day-night terminator over the course of several Gyrs. Whether or not this could be the source that ultimately causes a signal in the thermal phase curve observations (and thereby explain the hot spot shift observed by \citet{Demory2016a}) is, however, difficult to say. As soon as the plume interacts with the magma ocean, the temperature inside the magma ocean will equilibrate very quickly on a timescale that is much faster than the dynamics of the plume. Also, the heat flux from the plume will be much smaller than the heat flux that is radiated into space by the dayside magma ocean (a few tens of mW/m$^{2}$ vs. $\approx 10^{6}$ W/m$^{2}$).
The super-plume will lose some of its heat before it reaches the magma ocean because heat diffuses into the surrounding mantle. Most importantly, however, if the super-plume is able to sample a deep reservoir of 55 Cancri e's mantle, it will transport this material towards the upper layer of the mantle and into the magma ocean, from which it could be outgassed into the atmosphere. For the models in which the nightside is not fully molten, the super-plume is preferentially on the dayside of the mantle, and this could therefore lead to a dayside atmosphere that is chemically distinct from the nightside. 
It has been proposed that SiO that gets outgassed from a magma ocean on the dayside could lead to heterogeneous formation of clouds between the dayside and nightside, possibly explaining the observed hot-spot shift \citep{Hammond2017}. SiO is expected to be the major Si-bearing gas in the atmospheres of hot volatile-free super-Earths \citep{Schaefer2009,Zilinskas2022,Wolf2023}, and is the most promising species for emission spectroscopy with JWST \citep{Zilinskas2023}. 
\citet{Schaefer2009} also show that titanium (Ti) and possibly iron (Fe) might be depleted on the dayside because their condensates might be deposited on the nightside, removing them from the magma ocean reservoir. This process is also called trans-atmospheric distillation \citep{Kite2016}, whereby volatile rock-forming components partition into the atmosphere on the dayside and are then transported towards the nightside by winds, where they condense. 
Our results show that a super-plume could act as mechanism to re-enrich the magma ocean with species that condensed on the nightside. Whether this process is efficient enough to create a dayside and nightside surface and atmospheric composition that is similar will depend on whether mass recycling between the solid interior and overlaying magma ocean is more efficient than atmospheric transport of volatiles towards the nightside.

For surface temperatures of $T_{\mathrm{day}}=2750$\,K and $T_{\mathrm{night}}=1100$\,K, the solid mantle on the nightside flows with a velocity $\approx 0.2-1$\,cm/year and the plume rises with $\approx 2-5$\,cm/year. Hence, material needs around $1.5$\,Gyrs to descend into the deep mantle and another $200$\,Myrs to rise towards the dayside surface. The timescale for material to be transported from the surface of the nightside towards the dayside surface is therefore on the order of magnitude of Gyrs. This is approximately an order of magnitude higher than for Earth, where the transport (overturn) time scale is on the order of $100$\,Myrs. 
Hence for 55 Cancri e, trans-atmospheric distillation is expected to be dominant compared to the exchange between the solid mantle and magma ocean \citep{Kite2016}. 
It is therefore unlikely that the super-plume is able to replenish the dayside surface with material from the nightside if it requires Gyrs to move material from the nightside towards the dayside.
If heat transport in the magma ocean is efficient even if it is sub-adiabatic, the magma ocean is thicker on the dayside and the velocities are larger because of the higher mantle temperature and lower viscosities. In this case, the transport timescale would only be $\approx 10$\,Myrs and the magma ocean could therefore be efficiently replenished with material from the nightside. However, these models were only run for $1-4$ transport time scales due to computational resource limits. The models would need to be run for longer to see if the mantle reaches it steady state at this high temperature or if it will be able to cool down more, which would increase viscosity and decrease the mantle velocities.

If the nightside is molten, material can also be transported through lateral flow from the nightside towards the dayside \citep{Nguyen2020}. In our models, we find that the super-plume does not have a preferred location if the magma ocean covers the whole planet. Therefore, the composition of the atmosphere would be mostly dictated by magma ocean and atmospheric dynamics, because the material that gets advected by the super-plume into the magma ocean will be transported towards the dayside by the magma ocean on timescales that are much faster than the solid dynamics \citep[e.g.,][]{Boukare2022}. Further studies also suggest that the temperature gradient between the dayside and nightside can drive horizontal advection from the dayside towards the nightside \citep[e.g.][]{Hughes2007, Kite2016, Boukare2022}. Here, we assume that horizontal heat transport in the magma ocean is still very efficient even if it is sub-adiabatic. It is however difficult to say whether this is physically realistic. \citet{Kite2016} have shown that horizontal heat transport in a magma ocean can be much lower than the heat provided from stellar irradiation. Additionally, most studies that studied horizontal convection \citep{Rossby1965} have focused on cases of differential heating at the lower boundary \citep[e.g.][]{Mullarney2004, Hughes2007, SanmiguelVila2016}. Further research is therefore needed to investigate horizontal convection for laterally varying surface heating. A future study could also investigate how far the magma ocean extends towards the nightside for different efficiencies of horizontal heat transport.

\subsection{Magma Ocean thickness} \label{sec:discussion_thickness}
Our models show that the magma ocean's thickness on the dayside is only a small fraction of the whole mantle depth ($<7\%$) because vertical heat transport is suppressed if the magma ocean is sub-adiabatic. The magma ocean thickness on the dayside is around $500$\,km and $150$ to $500$\,km on the nightside (although for the models with a nightside temperature of $1100$\,K and dayside temperature of $2600$\,K it is only partially molten on that side) (Fig.~\ref{fig:mo_depth_allmodels}).
The dayside magma ocean is thicker if vertical heat transport is enhanced regardless of the temperature gradient inside the magma ocean. In that case, the magma ocean has a thickness of roughly $2000$\,km to $3000$\,km. For highly irradiated planets, if vertical heat transport in the magma ocean is efficient regardless of the temperature gradient, the magma ocean thickens due to its enhanced ability to transport heat from the star deeper into the mantle. This stands in contrast to terrestrial planets in magma ocean states where stellar irradiation plays a negligible role as a heat source, causing the magma ocean to thin more rapidly as the convective vigour (and therefore the heat transport efficiency) increases \citep[e.g.][]{Abe1997, Elkins-Tanton2012}. It is however uncertain how likely efficient vertical heat transport is in a magma ocean that is intensely heated from the top, and `adiabat-mode' is therefore physically more realistic. 

K2-141b is a super-Earth ($1.51R_{\oplus}$, $5.08 M_{\oplus}$, \citep{Malavolta2018}) that also falls into the category of lava planets. The thermal phase curve observation suggests a day-side temperature around $2050$ K and there is no indication of thermal emission from the nightside \citep{Zieba2022}. \citet{Boukare2022} used thermochemical models to show that the magma ocean depth of this lava planet could extend down to the core-mantle boundary. In our models, we do not observe such deep magma oceans because the magma ocean is mostly sub-adiabatic and therefore it would not be vigorously convecting. 
However, there might be other ways to transport heat efficiently in the magma ocean, which are not included in our models. 55 Cancri e has an orbital period of around $18$\,hours, resulting in a Rossby number---the ratio of inertial forces to Coriolis force---of $R_o \approx 0.001$.  This indicates that the Coriolis force could play an important role. The dynamics of a magma ocean at very low Rossby number are dominated by a strong shear flow \citep{Moeller2013}, which would prevent efficient heat transport through the magma ocean. However, a strong shear flow can also lead to Kelvin-Helmholtz instabilities \citep{Baines1994}, which could provide a mechanism to mix the magma ocean, similarly to Earth's oceans \citep{Smyth2012}.   

We therefore run several models where heat transport in the magma ocean is very efficient regardless of the adiabatic gradient, even if the actual mechanism for efficient heat transport is unknown. For these cases we find that the magma ocean extends to very large depths (Fig.~\ref{fig:mo_depth_allmodels}), especially on the dayside. 
For the highest dayside surface temperature investigated, the magma ocean reaches approximately $3000$\,km, corresponding to $40\%$ of the mantle. 
It is almost $5$ times as deep on the dayside compared to the nightside, and the super-plume is now always preferentially on the dayside.
A magma ocean that extends all the way to the CMB is however unlikely without any other (external or internal) sources of heat.  A possible additional heat source is, for example, tidal dissipation \citep{Bolmont2013}. Heat dissipation due to induction heating can most likely be excluded as the host star 55 Cancri is a slow rotator with a period around $39$\,days \citep{Bourrier2018, Folsom2020} and therefore most likely does not produce a strong enough magnetic field to cause significant induction heating \citep{Kislyakova2017, Kislyakova2020}.

The depth of the magma ocean is also influenced by the choice of the solidus and liquidus curves (Fig.~\ref{fig:solidus_liquidus}A). However, uncertainties regarding the melting properties of the mantle, particularly at greater depths, still persist \citep[e.g.][]{Andrault2017}. In this study, we employed the solidus curve from \citet{Zerr1998}. More recent experiments have since also yielded measurements of solidus and liquidus curves for mantle material under high pressure \citep[e.g.][]{Fiquet2010, Andrault2011, Nomura2014, Andrault2018, Pierru2022}, as well as more updated data from ab-initio and thermodynamic calculations \citep[e.g.][]{Liebske2012, Koker2013, Boukare2015, Miyazaki2019a} which could be considered in future studies  . At $136$\,GPa (pressure at Earth's CMB) the solidus in \citet{Andrault2011} reaches around $4140$\,K which is around $1160$\,K higher than a lower estimate provided by \citet{Boukare2015}. The liquidus at $136$\,GPa reaches around $5220$\,K in \citet{Fiquet2010}, whereas it is around $3900$\,K in \citet{Boukare2015}. A comparison between the solidus/liquidus curves at higher pressures used in this study and from more recent high-pressure experiments and thermodynamic calculations is shown in Figure~\ref{fig:solidus_liquidus}B. Our choice of solidus falls within the range between the lower and higher estimates derived from more recent studies, while the liquidus curve might be slightly underestimated. Correcting for this could lead to larger extent of partial melt. A future study could therefore investigate the consequences of higher or lower solidus/liquidus temperatures on the interior dynamics. Nevertheless, it is unlikely that the global interior dynamics would change drastically because of this since the magma ocean depth is primarily governed by the efficiency of heat transport, rather than the specific solidus/liquidus curves. At shallower depths, however, the precise solidus/liquidus temperatures could play a crucial role in determining whether the nightside remains solid. 

In conclusion, more studies are needed on the fluid dynamics and thermodynamic properties of a magma ocean that is heated intensely from the top to assess the efficiency of heat transport and chemical mixing in a magma ocean. This is particularly important because the convective vigour of the magma ocean can influence both outgassing composition \citep{Lichtenberg2021} and efficiency \citep{Salvador2023}. 

\section{Conclusions} \label{sec:conc}
In this study, we used 2D spherical annulus interior convection models to investigate the possible interior dynamic regimes of super-Earth 55 Cancri e. The thermal phase curve observations suggest that the dayside and possibly even the nightside could harbour a magma ocean. We constrained the surface temperature using the results of general circulation models and modelled both the solid dynamics and parameterised melt dynamics within the mantle.

Our models show that if the dayside harbours a magma ocean and the nightside is solid or only partially molten, a degree-one convection pattern is established with a super-plume preferentially rising on the dayside and cold material descending into the mantle on the nightside. The return flow of material from the dayside towards the nightside is accommodated in the magma ocean.  
If both the dayside and nightside are molten, the (super-adiabatic) magma ocean will tend to equilibrate the temperatures at the interface between the magma ocean and the underlying solid mantle. Therefore, the solid mantle flow is unaffected by the surface temperature contrast and upwellings do not have a preferred location anymore. In some cases, a degree-2 pattern of flow can then also be stable over several Gyrs. 
In some cases, especially with lower surface temperatures, small downwellings can form in the upper mantle and bridgmanite layer. However, they are not strong enough to penetrate through the bridgmanite-pPv phase transition where they become more diffuse. Therefore, the global mantle convection regime is mostly dictated by the super-plume rather than downwellings. 
  
We find that the intense solar insolation from the star leads to a temperature profile inside the magma ocean that is mostly sub-adiabatic. In this case, the magma ocean is most likely not vigorously convecting and is not able to transport heat efficiently into the deep interior. In this case, the (dayside) magma ocean is rather thin ($\approx 500$\,km). For the models where vertical heat transport is efficient even if the temperature gradient is not super-adiabatic, the dayside magma ocean thickness can increase significantly ($\approx 2000$--$3000$\,km).    

If the dayside harbours a magma ocean, the composition of the atmosphere on that hemisphere would be dictated by outgassing from the magma ocean. Atmospheric circulation could transport volatiles from the dayside to the nightside where they could condense \citep{Kite2016, Nguyen2020}. This could lead to dayside and nightside atmospheres that are chemically distinct in the extreme end-member where atmospheric transport mixing is inefficient. Whether or not the rising plume is able to replenish the magma ocean with material from the nightside depends on the timescales of material transport between the hemispheres and how efficient mass gets recycled between the solid and molten interior.  

The super-plume is mostly located at the substellar point, but in some cases it can also move up to $30^{\circ}$ east or west of the substellar point. Although this is similar to the observed hot spot shift \citep{Demory2016a}, more models which focus on magma ocean interactions with such a super-plume are necessary to find out whether this shift could be explained by a super-plume. To further test this hypothesis, there is also the need for more thermal phase curve observations of super-Earths that are hot enough to harbour magma oceans. In particular lava worlds with a tenuous (or no) atmosphere are of great interest, as the impact of a super-plume on the hot spot shift would be more significant compared to super-Earths with atmospheres that are efficiently redistributing heat. 
\begin{acknowledgements}
T.G.M. and D.J.B. acknowledge SNSF Ambizione Grant 173992. T.G.M. was supported by the SNSF Postdoc Mobility Grant P500PT\_211044. Calculations were performed on UBELIX (\url{http://www.id.unibe.ch/hpc}), the HPC cluster at the University of Bern. T.L. was supported by the Branco Weiss Foundation, the Simons Foundation (grant no. 611576) and the SNSF (grant no. P2EZP2-178621). 
The authors are also grateful to the open source software which made this work possible: \textsc{matplotlib} \citep{Hunter2007}, \textsc{StagPy} \citep{Morison2022}, \textsc{numpy} \citep{Harris2020}, \textsc{SciPy} \citep{Virtanen2020}.
This research benefited from discussions and interactions within the framework of the National Center for Competence in Research (NCCR) PlanetS supported by the SNSF. The authors thank Brice-Olivier Demory for the useful feedback during discussions, and Gregor Golabek for helpful comments that improved the manuscript. We also thank Charles-Édouard Boukaré for taking the time to review this paper and giving valuable feedback, which greatly improved this work.  
\end{acknowledgements}

\onecolumn
\begin{appendix}
\section{Supplementary figures}
\begin{figure*}[h!]
    \centering
    \includegraphics[width=1.0\textwidth]{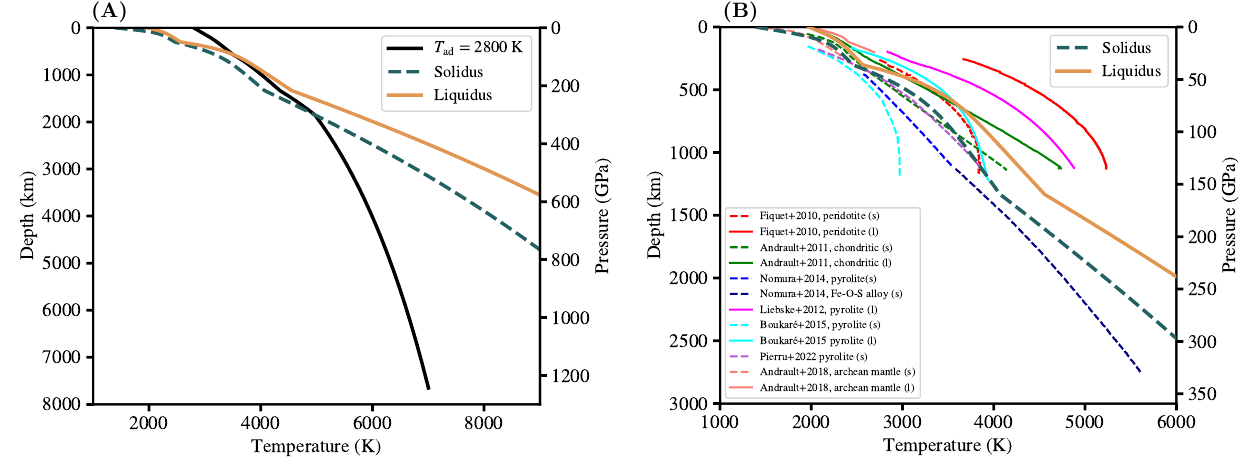}
    \caption{Solidus and liquidus curves. (A) shows the solidus and liquidus curves used in this study. The black line shows an adiabat with surface temperature $T_{\mathrm{surf}}=2800$\,K where latent heat effects are included in the mixed phase region ($0 < \phi < 1$). (B) shows a zoomed-in view of the region extending to a depth of 3000 km with comparison to different solidus and liquidus curves at higher pressures from experimental and thermodynamic calculation studies. (s) indicates the solidus curves (dashed lines) and (l) indicates the liquidus curves (solid lines).}
    \label{fig:solidus_liquidus}
\end{figure*}

\begin{figure*}
    \centering
    \includegraphics[width=0.8\textwidth]{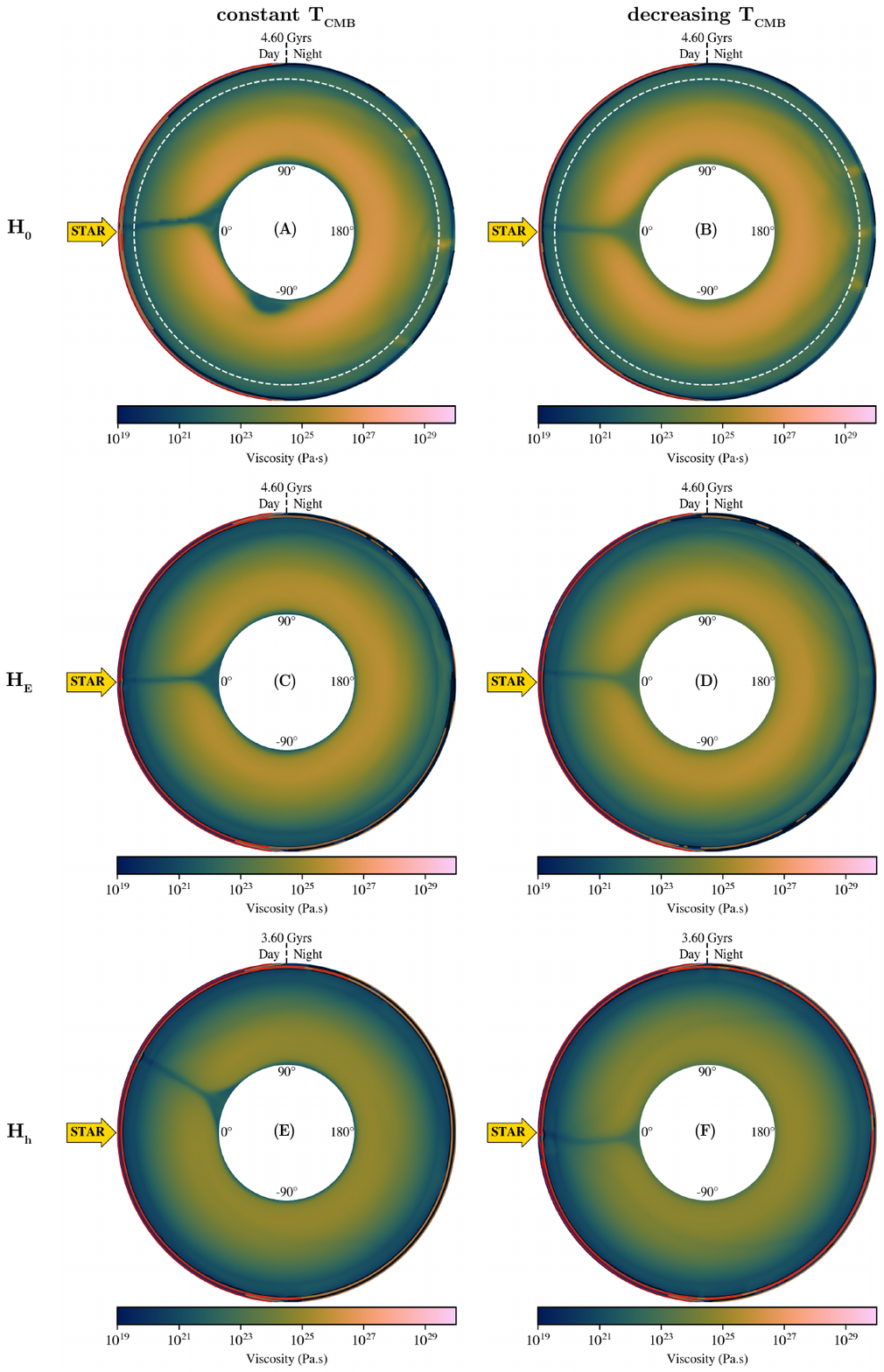}
    \caption{Snapshots of mantle viscosity for the 55 Cancri e models with $T_{\mathrm{day}}=2800$\,K and $T_{\mathrm{night}}=1150$\,K. The magma ocean is indicated in red (fully molten) and orange (partially molten). $H_{0}$, $H_{\mathrm{E}}$, and $H_{\mathrm{h}}$ indicate no, Earth-like, and high internal heating respectively. The left column shows the models with a constant CMB temperature. The right column shows the models where the CMB temperature decreases due to core cooling. Small downwellings can form in the bridgmanite layer for no internal heating, but they do not penetrate into the post-perovskite layer (indicated by the white dashed line). For higher internal heating, these downwellings become more diffuse.}
    \label{fig:cnc_ref_etamodels}
\end{figure*}

\begin{figure*}[h!]
    \centering
    \includegraphics[width=0.8\textwidth]{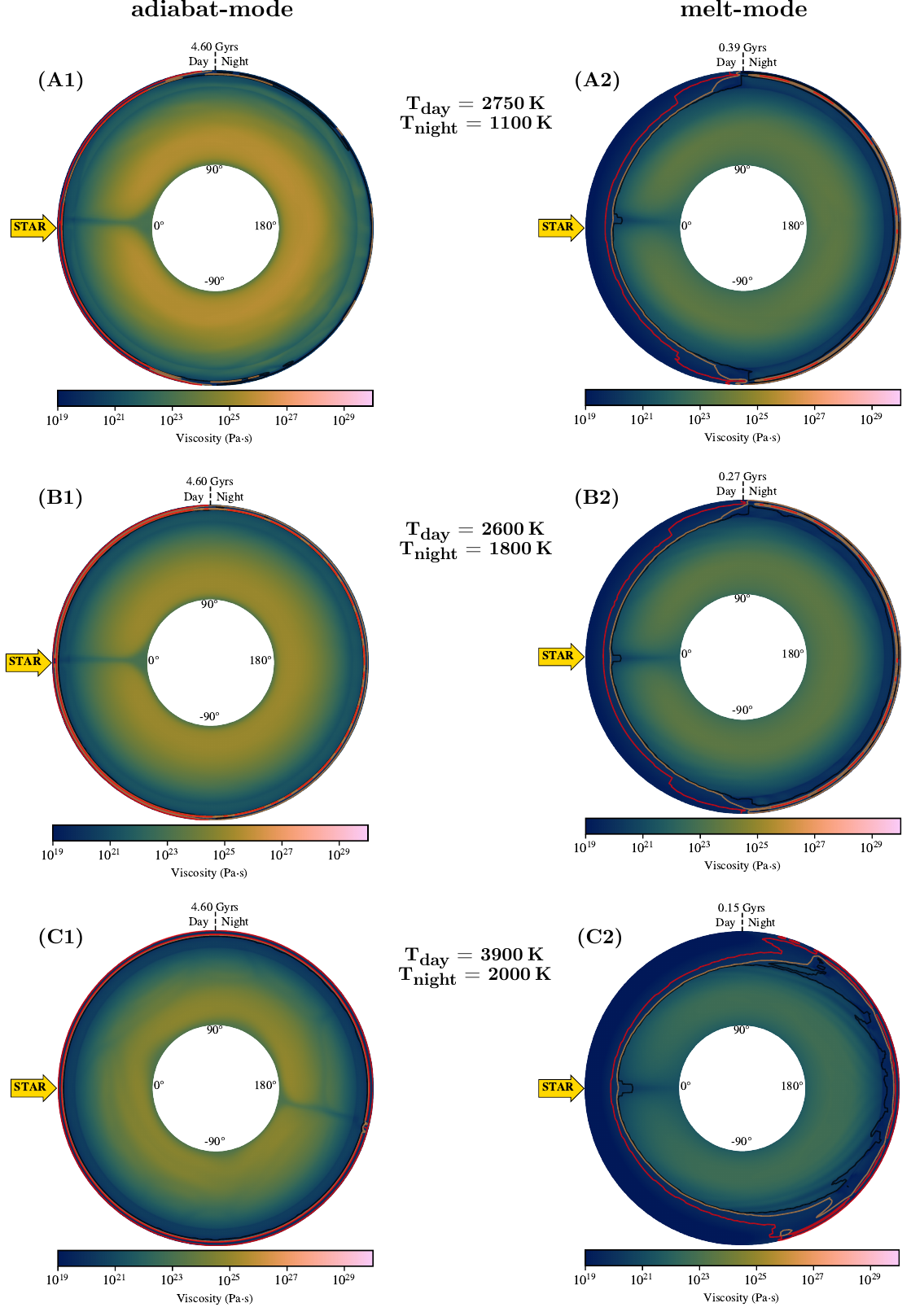}
    \caption{Snapshots of mantle viscosity for the `adiabat-mode' (left column) and `melt-mode' (right column) for different models of 55 Cancri e with different surface temperature contrasts. High velocities (larger than three times the mean mantle velocity) are indicated by red arrows. The magma ocean is indicated in red (fully molten) and orange (partially molten). If the nightside is not or only partially molten, an upwelling forms that is preferentially on the dayside (A,B). If both the dayside and nightside are fully molten, the upwelling does not have a preferred location anymore and can also be stable on the nightside (C).}
    \label{fig:cnc_tsurf_eta}
\end{figure*}

\end{appendix}	
\end{document}